# The Magnetically Induced Radial Velocity Variation of Gliese 341 and an Upper Limit to the Mass of Its Transiting Earth-sized Planet

Victoria DiTomasso[1,9], Mercedes López-Morales[1,2], Sarah Peacock[3,4], Luca Malavolta[5], James Kirk[6], Kevin B. Stevenson[7], Guangwei Fu[8], and Jacob Lustig-Yaeger[7]
[1] Center for Astrophysics | Harvard & Smithsonian, 60 Garden Street, Cambridge, MA 02138, USA; victoria.ditomasso@cfa.harvard.edu
[2] Space Telescope Science Institute, 3700 San Martin Drive, Baltimore, MD 21218, USA
[3] University of Maryland, Baltimore County, Baltimore, MD 21250, USA
[4] NASA Goddard Space Flight Center, Greenbelt, MD 20771, USA
[5] Dipartimento di Fisica e Astronomia "Galileo Galilei", Universitá di Padova, Vicolo del l'Osservatorio 3, I-35122 Padova, Italy
[6] Department of Physics, Imperial College London, Prince Consort Road, London, SW7 2AZ, UK
[7] Johns Hopkins APL, Laurel, MD 20723, USA
[8] Department of Physics and Astronomy, Johns Hopkins University, Baltimore, MD, USA

## Abstract

The Transiting Exoplanet Survey Satellite (TESS) mission identified a potential $0.88R_\oplus$ planet with a period of 7.577 days, orbiting the nearby M1V star GJ 341 (TOI 741.01). This system has already been observed by the James Webb Space Telescope (JWST) to search for presence of an atmosphere on this planet. Here, we present an in-depth analysis of the GJ 341 system using all available public data. We provide improved parameters for the host star, an updated value of the planet radius, and support the planetary nature of the object (now GJ 341 b). We use 57 HARPS radial velocities to model the magnetic cycle and activity of the host star, and constrain the mass of GJ 341 b to upper limits of 4.0 $M_\oplus$ ($3\sigma$) and 2.9 $M_\oplus$ ($1\sigma$). We also rule out the presence of additional companions with $M \sin i > 15.1$ $M_\oplus$, and $P < 1750$ days, and the presence of contaminating background objects during the TESS and JWST observations. These results provide key information to aid the interpretation of the recent JWST atmospheric observations and other future observations of this planet.

*Unified Astronomy Thesaurus concepts:* Planet hosting stars (1242); Exoplanet astronomy (486); Radial velocity (1332); Transit photometry (1709); Exoplanet detection methods (489)

## 1. Introduction

A major goal of exoplanet science is to find and characterize terrestrial planets ($R_p \lesssim 1.6R_\oplus$). Astronomers have primarily done this by searching for planets with radii and masses similar to Earth's, with more than 40 ($R_p < 1.6R_\oplus$, $M_p < 3M_\oplus$) now confirmed via transit and radial velocity detection.[10] A next step toward understanding these planets and assessing their properties is to study their atmospheres. Prior to the launch of the James Webb Space Telescope (JWST), it had only been possible to place strong constraints on the atmospheric properties of a small number of terrestrial planets: e.g., LHS 3844 b (L. Kreidberg et al. 2019) and the TRAPPIST-1 planets (e.g., J. de Wit et al. 2016, 2018; S. E. Moran et al. 2018; H. R. Wakeford et al. 2019; A. Gressier et al. 2022; L. J. Garcia et al. 2022). Since its launch in 2021 December, JWST has spent significant resources to definitively detect and characterize rocky exoplanet atmospheres. The first wave of publications include: LHS 475 b, which has either a high-altitude cloud deck or a thin secondary atmosphere (J. Lustig-Yaeger et al. 2023); GJ 486 b, which was found to have water in its transmission spectra coming from either the atmosphere of the planet or cool spots on the surface of the star (S. E. Moran et al. 2023); GJ 1132 b, with two observed transmission spectra that yield inconsistent results (E. M. May et al. 2023); and planets b and c in the TRAPPIST-1 system, which both appear to have little or no atmosphere (T. P. Greene et al. 2023; S. Zieba et al. 2023).

LHS 475 b, GJ 486 b, and GJ 1132 b in this list are part of a JWST Cycle 1 program (JWST-GO-1981; PIs: Stevenson and Lustig-Yaeger) that is observing five warm terrestrial planets around nearby M dwarfs to establish if they have atmospheres. The other two targets in that program are TRAPPIST-1h and TOI-741.01. TOI-741.01 was cataloged by the Transiting Exoplanet Survey Satellite mission (TESS; G. R. Ricker et al. 2014) as a potential 0.88 $R_\oplus$ planet orbiting the nearby M dwarf GJ 341 with an orbital period of 7.577 days. However, a follow-up analysis based on the TESS observations alone concluded that this object was neither likely to be a planet nor a false positive (S. Giacalone et al. 2020), and an analysis using publicly available TESS and the High Accuracy Radial velocity Planet Searcher spectrograph (HARPS; M. Mayor et al. 2003) data, which we use in this work, assigned it a 92.6% false-alarm probability (S. Palatnick et al. 2021). In spite of these uncertainties, JWST observed TOI-741.01 over three predicted transit epochs between 2023 March 2 and April 17 UT.

In this paper, we demonstrate the planetary nature of TOI-741.01 (hereafter referred to as GJ 341 b), by analyzing all public existing data for this target and its host star. In addition, we provide improved planetary and stellar parameters to facilitate future atmospheric characterization of this planet. Section 2 describes the data used in this study, and Section 3 characterizes the host star, including a presentation of its stellar parameters, magnetic cycle, and stellar activity. Section 4 describes the analysis of the planetary signals present in the TESS and JWST light curves (LCs), while Section 5 describes the analysis of the HARPS radial velocity (RV) data and how we arrive at the upper-limit mass measurements of GJ 341 b. Results and discussion are given in Section 6.

[9] National Science Foundation Graduate Research Fellow.
[10] https://exoplanetarchive.ipac.caltech.edu/

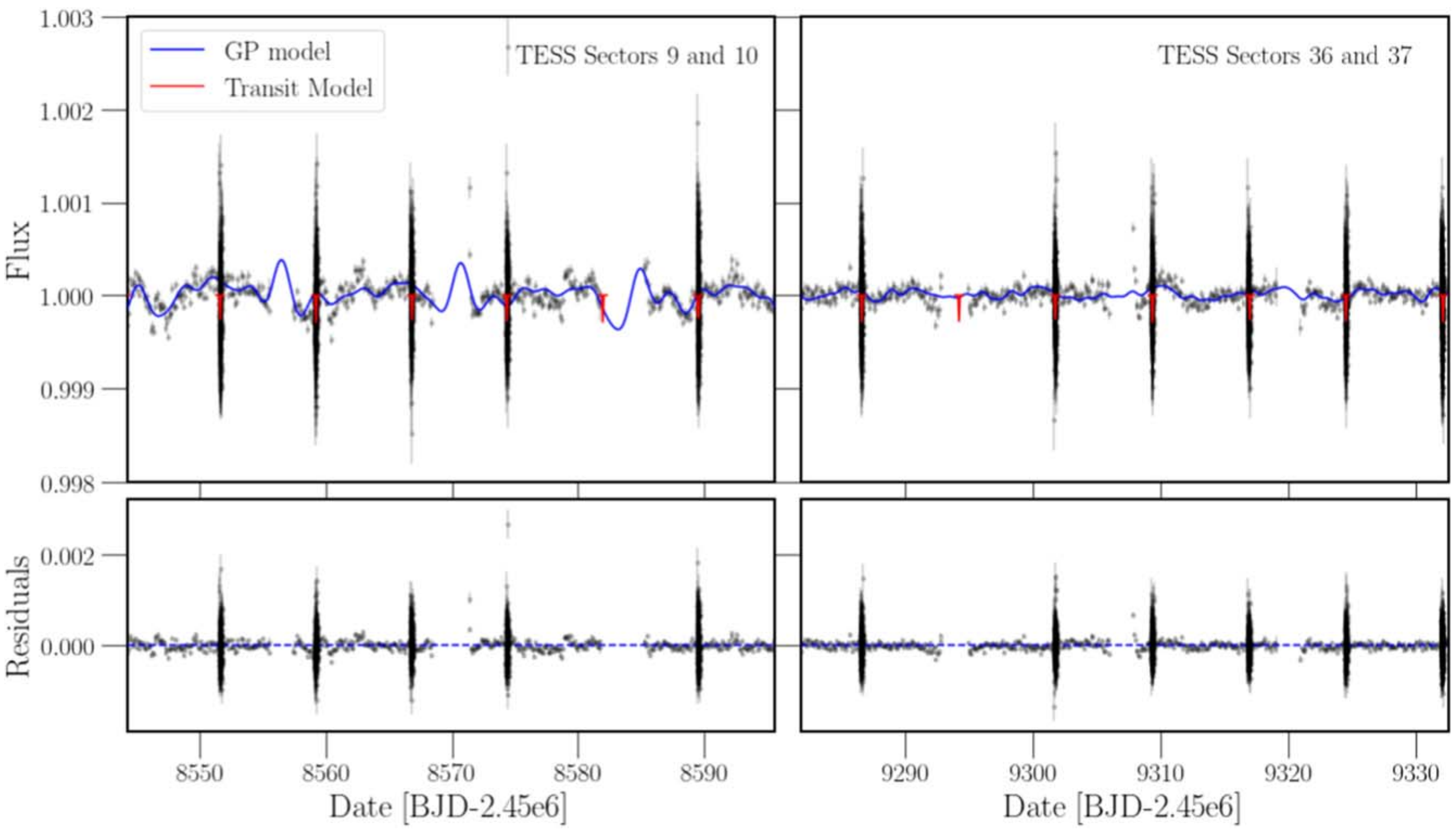

**Figure 1.** TESS light curves for GJ 341, which was observed during sectors 9 and 10 (left) and 36 and 37 (right). The in-transit times appear darker, with a larger spread in flux values, because we used every point of the 2 minute cadence data when fitting our transit model (red). The phase-folded transits are shown in Figure 3. When fitting our GP model (blue) to the out of transit data, we binned the data down to a cadence of one point per 2 hr, as represented in this figure. We fit two separate GP models, one to sectors 9 and 10 and the other to sectors 36 and 37. The residuals of the best-fit GP and keplerian models are plotted in the lower panels.

## 2. Data

### 2.1. TESS Photometry

TESS observed GJ 341 with a 2 minute cadence in sectors 9 and 10 (2019 February 28–April 22), and with a 2 minute and 20 s cadence in sectors 36 and 37 (2021 March 7–April 28). For our analysis, we used the Pre-search Data Conditioning Simple Aperture Photometry (J. C. Smith et al. 2012; M. C. Stumpe et al. 2012) flux and flux uncertainties as reduced by the Science Processing Operations Center pipeline and provided on the Mikulski Archive for Space Telescopes (MAST).[11] We removed all points with NaN flux values or bad-quality flags (DQUALITY > 0). We then normalized each sector by dividing the light curve by the mean sigma-clipped, out-of-transit (OOT) flux value. We opted to use only the 2 minute cadence photometry since it was sufficient to characterize the planet's transit and helped reduce the computational time to fit our transit and stellar activity models.

We divided the TESS LCs into in-transit and OOT data sets (used in Sections 4.1 and 3.3.1, respectively), where in-transit was defined as within 3 hr of the central transit time. In doing so, we were able to vary the cadence of the data that we fit for the two models—binning the TESS OOT photometry down to a cadence of one point per 2 hr, to fit for longer-term effects of stellar spots and rotation, while using the full 2 minute cadence data to precisely fit the transit parameters. The full TESS data set, as we used it, is plotted in Figure 1.

### 2.2. ASAS-SN Photometry

We also used $g$-filter aperture photometry from the All-Sky Automated Survey for Supernovae (ASAS-SN) to characterize the star's activity (B. J. Shappee et al. 2014; C. S. Kochanek et al. 2017). ASAS-SN monitored GJ 341 from 2018 June until 2023 July, with a usual cadence of three points per night of observation and a median flux uncertainty of 1.742 mJy. We retrieved these data using ASAS-SN Sky Patrol,[12] and similar to our treatment of the TESS OOT light curves, we binned the ASAS-SN data to one point per 12 hr. These data are plotted in Figure B1.

### 2.3. HARPS Spectroscopy

We used 57 publicly available RV measurements collected with the HARPS spectrograph, mounted at the ESO 3.6 m telescope in La Silla (M. Mayor et al. 2003) between 2003 December and 2013 January, with a median uncertainty of 1.07 m s$^{-1}$, and published in T. Trifonov et al. (2020).[13] Specifically, we used the *DRVmlcnzp* RVs as described in T. Trifonov et al. (2020). We used the stellar activity indicator measurements derived in T. Trifonov et al. (2020) using the SpEctrum Radial Velocity AnaLyser (M. Zechmeister et al. 2018) pipeline. The RV time series are plotted in Figure 2, and the activity indicators are plotted versus the RVs in Figure B2.

[11] All the TESS data used in this paper can be found on MAST: doi:10.17909/65zq-g239.

[12] https://asas-sn.osu.edu/sky-patrol/coordinate/5f015161-e226-4d4e-9a0e-7302ae0750ac

[13] https://cdsarc.cds.unistra.fr/viz-bin/cat/J/A+A/636/A74

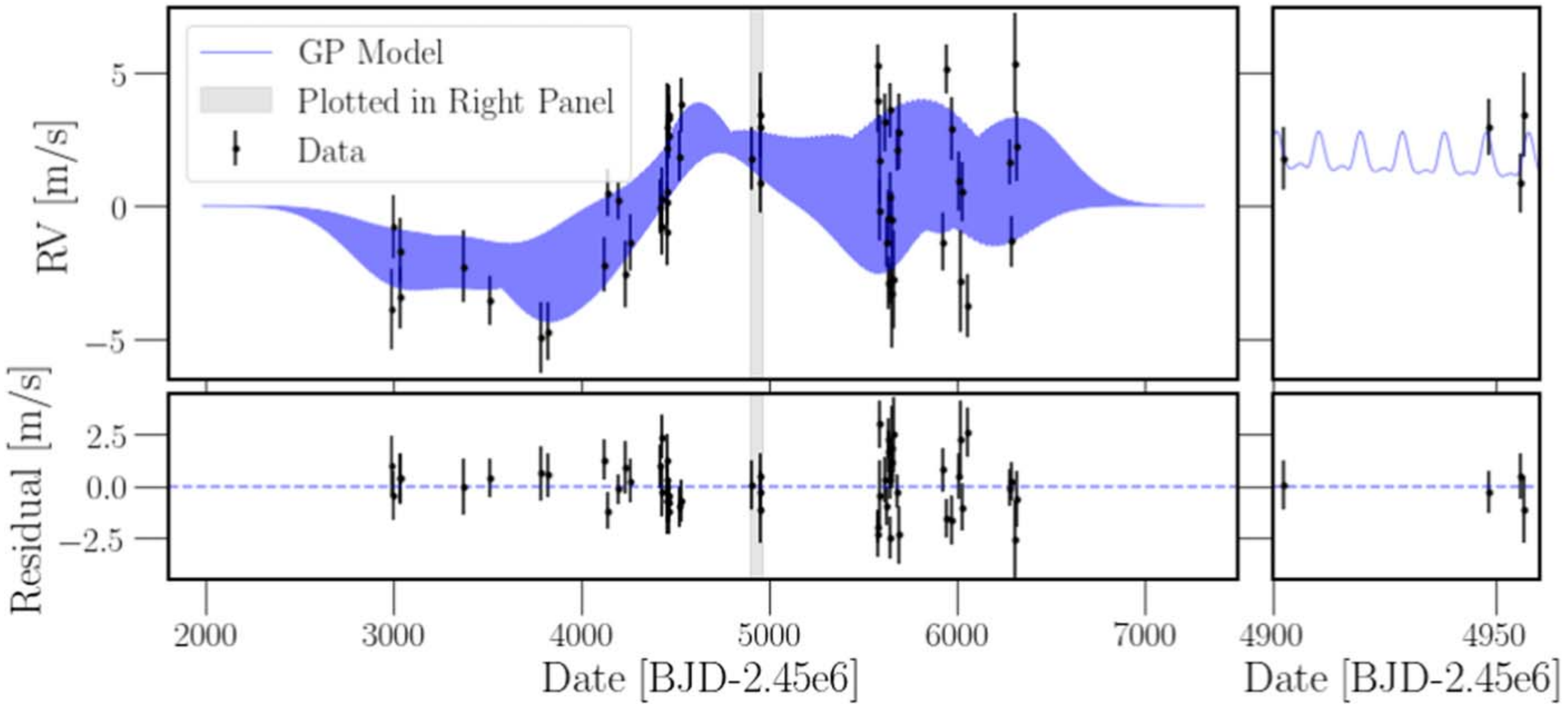

**Figure 2.** GP model fit to the RV data. Left panel shows the entire RV time series in black and the GP model in blue, with residuals plotted in the lower panel, to demonstrate that a quasiperiodic GP model was able to capture the long-period variation caused by the stellar magnetic cycle. The gray shaded region is shown in the right panel. The right panel shows the same model and data over a centrally located 60 day baseline, to highlight the effects of the 9.42 day stellar rotation period.

### 2.4. JWST White Light Curves

Finally, we gathered data from three transit observations of GJ 341 b with JWST, collected on 2023 March 2, March 10, and April 17 UT. The three transits were observed with JWST's Near Infrared Camera (NIRCam; M. J. Rieke et al. 2023), as part of GO program 1981, using the F444W filter and the SUB-GRISM128 subarray, which yield spectra in the wavelength range between 3.8 and 5.1 $\mu$m at a native resolution of $R$ ~ 1650. Each data set consists of a time series of 4652 integrations, with three groups per integration, with effective integration times of 3.38 s.

We reduced the data with three independent pipelines: `Eureka!` (T. Bell et al. 2022), `Tiberius` (J. Kirk et al. 2017, 2021), and `tswift`, as described in more detail in J. Kirk et al. (2024). In brief, each pipeline started by running the `uncal.fits` files through different modified versions of Stage 1 of the `jwst` pipeline. The three pipelines then independently performed outlier detection, traced the stellar spectra, and performed background subtraction and aperture photometry. At each step, each pipeline used slightly different parameters (e.g., aperture size and standard vs. optimal extraction). The full set of parameters and assumptions made by the reduction are detailed in J. Kirk et al. (2024). The result of each pipeline's reduction that is used in this analysis are the white light curves, which were integrated over 3.90–5.00 $\mu$m for `Eureka!`, 3.89–5.02 $\mu$m for `Tiberius`, and 3.92–4.99 $\mu$m for `tswift`. Each pipeline then performed an independent fit to its white light curves. The phase-folded JWST white light curves are plotted in Figure 3 (`Eureka!`) and Figure B3 (`Tiberius` and `tswift`).

## 3. Stellar Characterization

GJ 341 is an M1 star, just over 10 pc from Earth, which has long been studied as a nearby star (e.g., T. W. Russo 1956; V. A. Zakhozhaj 1979), a high-proper-motion star (e.g., W. J. Luyten 1979), and a photometric and RV standard star (e.g., D. Kilkenny et al. 2007; C. Soubiran et al. 2018; A. Antoniadis-Karnavas et al. 2020). It was also identified as a favorable candidate for exoplanet studies, e.g., for detecting a rocky planet in the habitable zone via RV monitoring (S. Hojjatpanah et al. 2019) and for obtaining spectroscopy of habitable planets via nulling interferometry (A. Léger et al. 2015). This star has also been labeled as a potential spectroscopic binary (A. Reiners et al. 2012). To this wealth of literature, we add stellar characterization via spectral energy distribution (SED) analysis, as well as an analysis of the star's magnetic cycle, possible binarity and stellar activity signatures. We refer to the star's years- or decades-long magnetic activity cycles, equivalent to the Sun's 11 yr magnetic cycle, as its "magnetic cycle." We refer to the star's variation as caused by its rotation and spot decay timescale as "stellar activity."

### 3.1. Stellar Parameters

Spectroscopic analysis and SED fitting are two primary methods to determine the stellar parameters of M dwarfs. The same HARPS spectra used to derive the RVs in this work have already been used to determine GJ 341's effective temperature ($T_{\rm eff}$) and metallicity (Fe/H) by J. Maldonado et al. (2020), and so we opt to perform an SED fit to determine these and other stellar parameters. We list the spectroscopic measurements alongside our SED results in Table 1 and find them to be consistent.

We constrained the effective temperature ($T_{\rm eff}$), surface gravity ($g$), mass, and radius of GJ 341 using a tailored grid of `PHOENIX` models guided by 116 photometric fluxes from the Two Micron All Sky Survey (2MASS) $J$, $H$, $K$, Johnson $V$, $J$, $H$, $K$, Gaia $G$, and Cousins $I$, $R$ bands as downloaded from the Vizier Photometric Viewer.[14] Expanding around preexisting estimates for the parameters, we computed 120 models with $T_{\rm eff}$ = 3700–3900 K, $\Delta T$ = 10 K; $\log(g)$ = 4.71–4.73 (cgs), $\Delta\log(g)$ = 0.01 (cgs); $M_\star$ = 0.45–0.55 $M_\odot$, $\Delta M_\star$ = 0.1 $M_\odot$ (M. C. Turnbull 2015; K. G. Stassun et al. 2019), and

[14] http://vizier.cds.unistra.fr/vizier/sed/

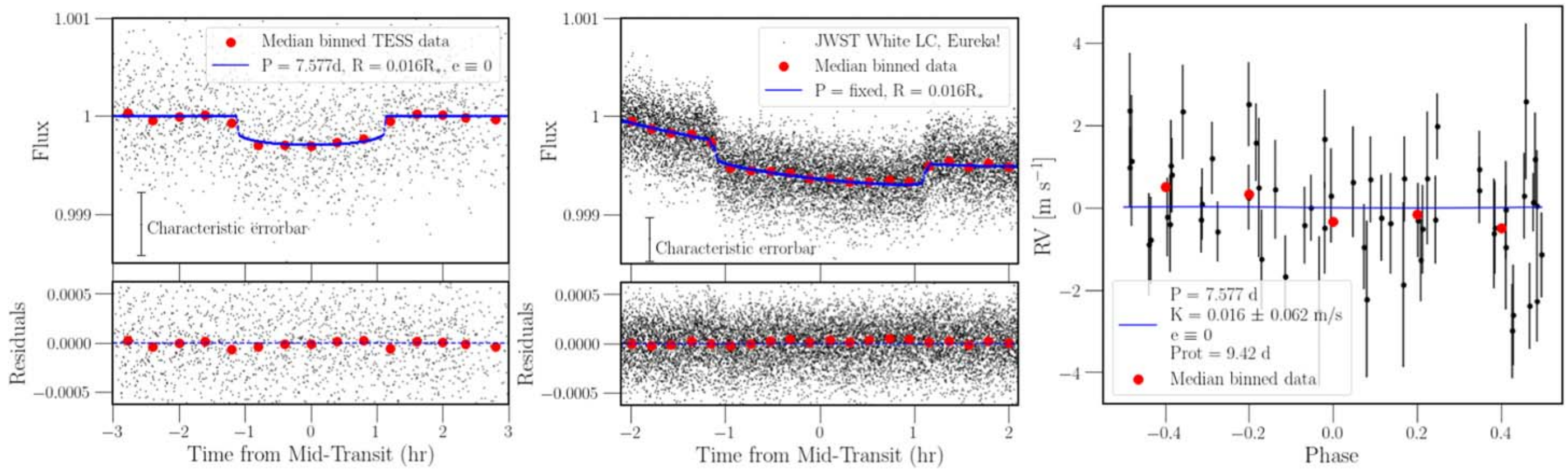

**Figure 3.** Phase-folded fit of the TESS transits (left), JWST transits (middle), and the HARPS RVs (right) for the 7.577 day planet. The left panel shows the phase-folded TESS light curve in black, including a characteristic error bar, the circular orbit transit model in blue, and the median data value within a bin of 0.4 hr in red. The upper portion shows the model plotted over the data and the lower portion shows the residuals of the data minus the model. The middle panel shows the same for the JWST white light curve, with the red dots showing the median binned data for bins of 0.2 hr. The right panel shows the phase-folded RV fit for the 7.577 day planet in the HARPS RVs, once the GP has been subtracted. Again, the data are plotted in black, the model in blue, and the median data value within a bin of 0.2 phase in red. Since we were not able to detect the planet in these RVs, the Keplerian model is a flat line at 0 m s$^{-1}$. We omit a residual panel, as the model is equal to zero.

**Table 1**
GJ 341 Stellar Parameters

| Stellar Properties | Value |
|---|---|
| Alternate names | HD 304636 |
| | HIP 45908 |
| | TIC 359271092 |
| | TOI-741 |
| R.A. (deg) | 140.39914047698 [b] |
| Decl. (deg) | −60.28114385834 [b] |
| *V*mag | 9.465 [c] |
| $T_{\rm eff}$ (K) | 3770 ± 40 [a] |
| | 3798 ± 69 [d] |
| [Fe/H] (dex) | −0.16 ± 0.09 [d] |
| log(*g*) (cgs) | 4.72 ± 0.02 |
| $M_\star$ ($M_\odot$) | 0.48 ± 0.03 |
| $R_\star$ ($R_\odot$) | $0.5066^{+0.0169}_{-0.0172}$ |
| $L_{\rm bol}$ ($10^{32}$ erg s$^{-1}$) | 1.76 ± 0.06 |
| SpT | M1 [d] |
| $v \sin i_\star$ (km s$^{-1}$) | 2.32 ± 0.65 [e] |
| Distance (pc) | 10.4495 ± 0.0015 [f] |

**Notes.** Values are from this work, unless otherwise noted.
[a] Adopted.
[b] Gaia Collaboration et al. (2023).
[c] C. Koen et al. (2010).
[d] J. Maldonado et al. (2020).
[e] S. Hojjatpanah et al. (2019).
[f] A. G. A. Brown et al. (2021).

computed synthetic photometry for each model in visible and near-infrared bands consistent with the available archival measurements. To isolate spectral shape, we normalized the models and observations to their corresponding 2MASS *J*-band flux density values before using a reduced $\chi sq$ test to identify the model that most closely reproduced the observations. The fully explored grid yielded $\chi^2_\nu$ values between 1.14 and 6.88, with 40 models returning similar values less than 1.5 (see Figure 4). We then used $R_\star^2/d^2$ with the measured Gaia EDR3 distance to scale the models such that $F_{\rm 2JMASS,mod} = F_{\rm 2JMASS,obs}$ to determine a stellar radius. These best-fit parameters are consistent with literature values, namely those from J. Maldonado et al. (2020), and are listed in Table 1.

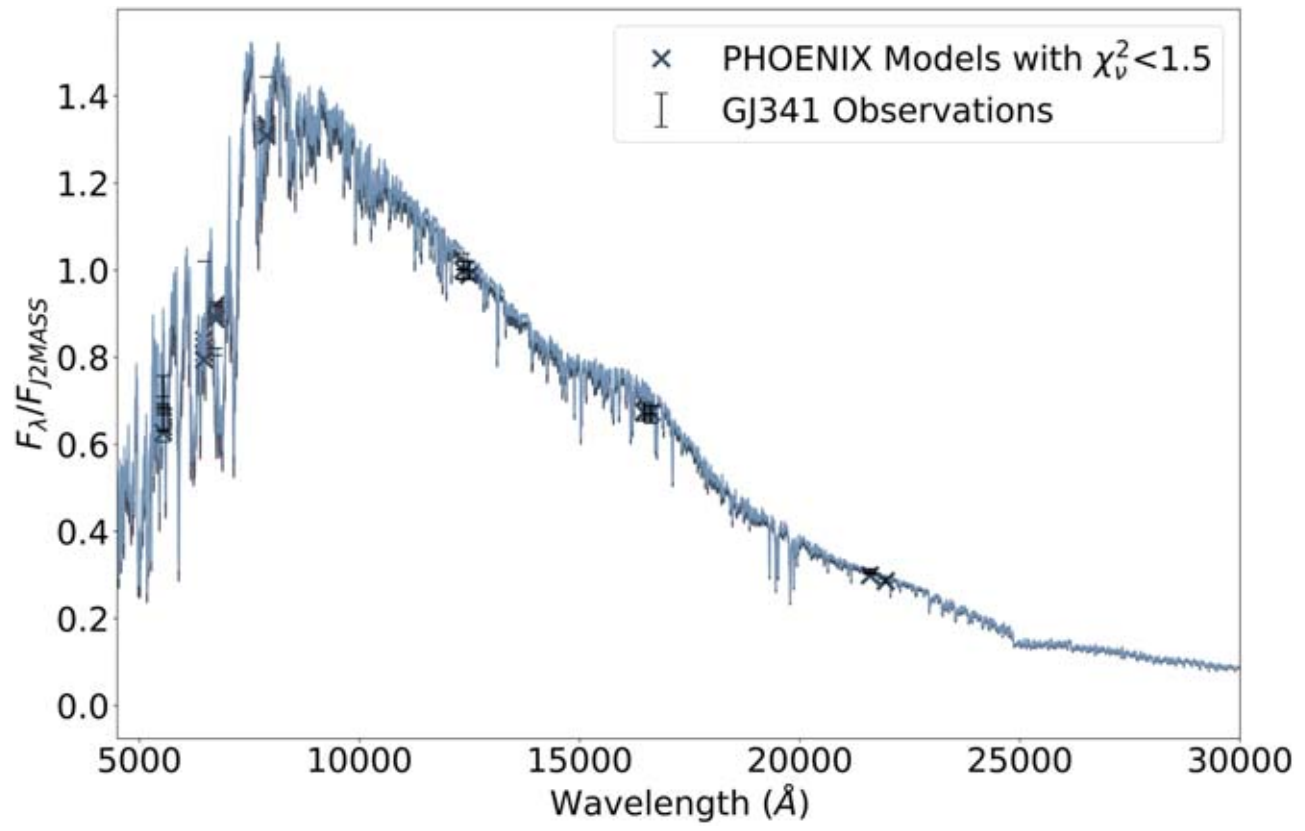

**Figure 4.** Comparison of the 40 closest matching ($\chi^2_\nu < 1.5$) PHOENIX models to archival GJ 341 photometry. These models have $T_{\rm eff} = 3730$–3810 K, log(*g*) = 4.70–4.74 (cgs), $M_\star = 0.45$–0.51 $M_\odot$.

### 3.2. Stellar Magnetic Cycle and Possible Binarity

In their early M star rotation and activity catalog, A. Reiners et al. (2012) listed GJ 341 as a possible spectroscopic binary. GJ 341 exhibits sine-like, long-timescale variation with an amplitude of ∼2.5 m s$^{-1}$ and a period of ∼4000 days. If this long-timescale variation were caused by a body orbiting GJ 341 (in addition to the transiting planet with a period of 7.577 days), it would correspond to a companion with a minimum mass of 40 $M_\oplus$. However, we find that this variation can be explained as evidence of the star's magnetic cycle, and present additional evidence that GJ 341 is not a binary.

We find that this sine-like variation also appears in several of the activity indicators, suggesting that this signal is caused by stellar activity and not by a companion. Na II D is the activity indicator that is most highly correlated with the RVs (with $r = 0.61$, as shown in Figure B2). In addition to being the most highly correlated, J. Gomes da Silva et al. (2011), a study of magnetic cycles on early M dwarfs, finds that the Na doublet tracks chromospheric activity in early M-type stars well, and recommends the use of the Na activity indicators for modeling the magnetic cycle of these stars. For these reasons, we chose to fit our magnetic cycle/stellar activity model to Na II D along with the RVs.

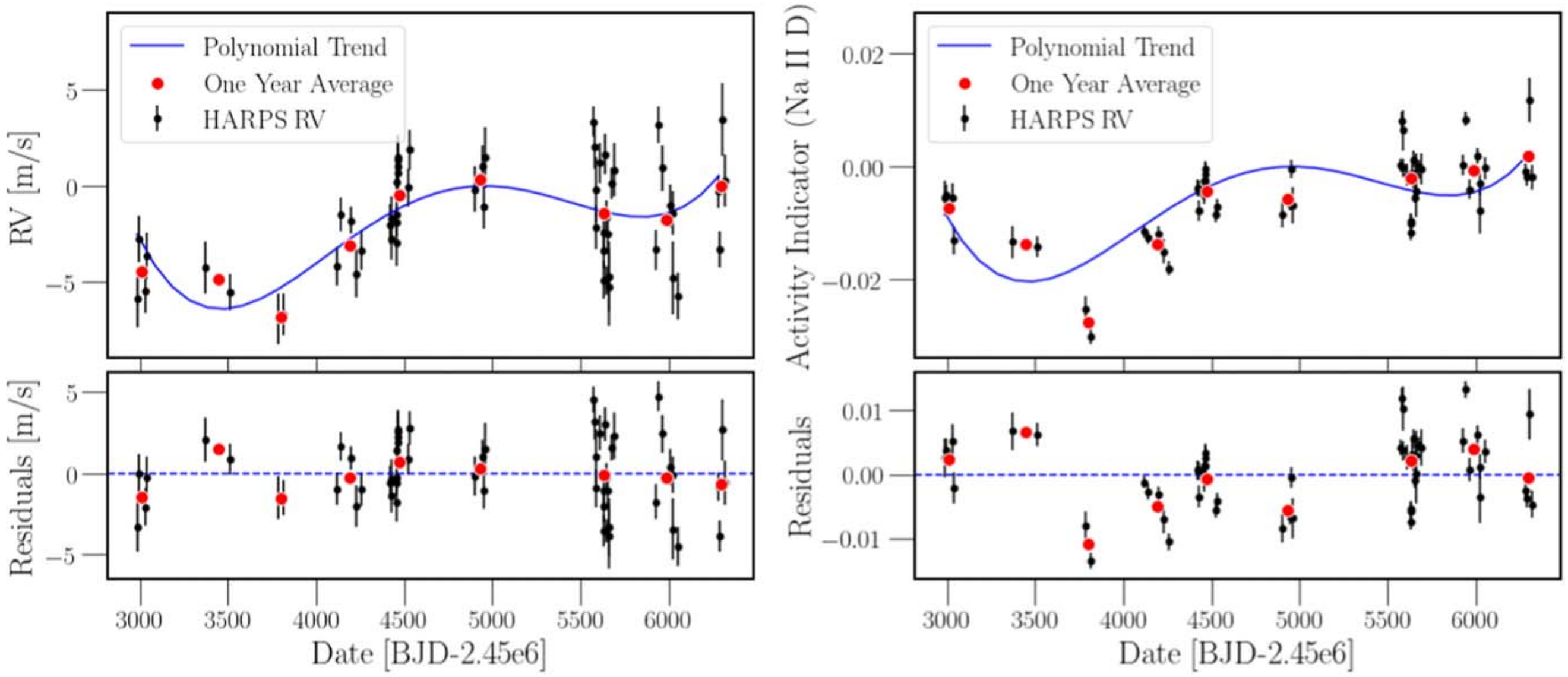

**Figure 5.** RV time series (left) and time series of stellar activity indicator Na II D (right) plotted with their simultaneously fit polynomial trend, as well as binned averages for each year of observations. There is a scale factor fit for each data set separately to account for the variation in amplitude between the two data sets. We tried subsequently fitting the residuals here with a GP to model stellar activity and a Keplerian to model the planet, and found this combination of models (polynomial plus GP and Keplerian) to be disfavored when compared to a GP and Keplerian without a polynomial model.

J. Gomes da Silva et al. (2011) fit the magnetic cycle of the stars in their sample with a fourth-order polynomial, fitting it first to the activity indicator time series, then scaling that trend to the RVs before subtracting it off of their RV data sets. We present the results of fitting GJ 341's RVs and Na II D activity indicator simultaneously with a fourth-order polynomial in Figure 5 to demonstrate that this trend is consistent with those found by J. Gomes da Silva et al. (2011). Ultimately, however, we opted to fit the RVs and Na II D with a multidimensional, quasiperiodic Gaussian process (GP), omitting a polynomial model, because the GP-only approach was able to fit the data equally well, even slightly reducing the scatter in the residuals, while reducing the number of free parameters.

As another test of binarity, we found that GJ 341's BP − RP color (2.023; Gaia Collaboration et al. 2023) and $V$ magnitude (9.465; C. Koen et al. 2010) align with those expected of a single star with a spectral type between M0.5V and M1.0V (M. J. Pecaut & E. E. Mamajek 2013).[15] Additionally, massive companions have been ruled out at separations of 5–70 au via high-resolution imaging with Hubble Space Telescope/NICMOS (S. B. Dieterich et al. 2012), at separations of 3–10,000 au via adaptive optics (K. Ward-Duong et al. 2015), and at separations of 1–1000 au via ground-based speckle imaging (K. V. Lester et al. 2021). The sensitivity curves reported in these studies are plotted as a function of angular separation in Figure B4. Finally, Gaia and Hipparcos astrometry do not indicate wider companions, and the Gaia Renormalized Unit Weight Error, which is expected to be 1.0 for single stars and indicates how well a target's astrometry is fit by a single-star model, is 1.021 for GJ 341 (P. Kervella et al. 2019; L. Lindegren et al. 2021; P. Kervella et al. 2022). We discuss the mass and period limits for detectable companions based on our RV analysis in Section 5.3.

### 3.3. Stellar Activity

We constrain GJ 341's stellar activity, i.e., photometric modulation caused by stellar rotation and stellar spots, by fitting the OOT TESS and ASAS-SN LCs and the RVs with GPs. Specifically we used a quasiperiodic kernel as implemented in `PyORBIT` (L. Malavolta 2016). This kernel has four hyperparameters: variability amplitude ($H_{\rm amp}$), nonperiodic characteristic length (associated with the spot decay timescale, $P_{\rm dec}$), variability period (associated with the stellar rotation period, $P_{\rm rot}$) and periodic characteristic length (associated with the number of spots/spot regions on the surface of the star, $w$). We placed priors on $P_{\rm rot}$ using the $v \sin i_\star$ measurement from S. Hojjatpanah et al. (2019) and the stellar radius, as derived in Section 3. This yielded $P_{\rm rot}$ bounds of 7.9–14.2 days, which we expanded to an upper limit of 15 days. We set an upper limit on $P_{\rm dec}$ of 1000 days to allow for long-lived starspots, and a prior of 0.5 +/- 0.05 on $w$ following S. V. Jeffers & C. U. Keller (2009).

#### 3.3.1. Light-curve Analysis

We fit quasiperiodic GPs to TESS sectors 9 and 10 separately from sectors 36 and 37 to avoid fitting the 2 yr gap between sectors 10 and 36, as well as to reduce computational time. We used `emcee` (D. Foreman-Mackey et al. 2013) with 20 walkers for the chains (4 times the number of dimensions in the model). We ran the sampler for 55,000 steps, excluded the first 25,000 as burn-in, and used a thinning factor of 100.

As shown in Figure B5, the GP fit large values for $P_{\rm dec}$ (sectors 9 and 10 resulted in a double-peaked posterior for $P_{\rm dec}$ with one around 250 days and the other around 600 days, and sectors 36 and 37 resulted in a single-peaked posterior at $702^{+128}_{-114}$ days), indicating that the stellar spots are stable over the timescale of at least two sectors, about 50 days. The fit $H_{\rm amp}$

[15] Accessed via the updated table at https://www.pas.rochester.edu/~emamajek/EEM_dwarf_UBVIJHK_colors_Teff.txt.

**Table 2**
GJ 341 b Planetary Parameters

| Planetary Properties | Value | Value | Value | Value |
|---|---|---|---|---|
| | TESS | JWST - Eureka! | JWST - Tiberius | JWST - tswift |
| $P$ (d) | $7.576860^{+0.000034}_{-0.000020}$ | Fixed to TESS value | Fixed to TESS value | Fixed to TESS value |
| $T_c$ (BJD-2.45e6) | 9301.771 ± 0.002 | $10006.420147^{+0.000152}_{-0.000141}$ | $10006.420152^{+0.000193}_{-0.000196}$ | $10006.420143^{+0.000241}_{-0.000245}$ |
| e | 0[a] | 0[a] | 0[a] | 0[a] |
| $a/R_\star$ | $24.5^{2.0}_{-4.0}$ | $20.12^{+2.93}_{-2.19}$ | $22.88^{+2.97}_{-4.37}$ | $23.48^{+1.44}_{-1.95}$ |
| $i$ (deg) | $89.24^{+0.53}_{-0.96}$ | $88.17^{+0.64}_{-0.51}$ | $88.77^{+0.88}_{-0.97}$ | $88.72^{+0.41}_{-0.46}$ |
| $R_p$ ($R_\star$) | 0.0160 ± 0.0008 | $0.016079^{+0.000271}_{-0.000254}$ | $0.015315^{+0.000316}_{-0.000297}$ | $0.016130^{+0.000295}_{-0.000337}$ |
| $R_p$ ($R_\oplus$) | 0.88 ± 0.05 | $0.89^{+0.05}_{-0.04}$ | 0.85 ± 0.05 | 0.89 ± 0.05 |
| $S_p$ ($S_\oplus$) | $13.8^{+5.9}_{-3.7}$ | $20.5^{+6.5}_{-8.0}$ | $15.8^{+7.6}_{-5.7}$ | $15.0^{+4.0}_{-3.4}$ |
| $M_p$ ($M_\oplus$) | $< 4.0$[b] | … | … | … |
| Transit Duration (hr) | $2.23^{+0.12}_{-0.08}$ | … | … | … |

**Notes.**
[a] Fixed.
[b] Upper limit derived from HARPS RVs.

and $w$ values were consistent between the two pairs of sectors. $P_{\rm rot}$, however, varied between data sets, with the $P_{\rm rot}$ posterior for sectors 9 and 10 showing two peaks at around 13.2 and 13.7 days, whereas sectors 36 and 37's posterior peaks at $14.55^{+0.13}_{-0.12}$ days. The best-fit GP models for both sector pairs are plotted in Figure 1.

We similarly fit the ASAS-SN light curves with a quasiperiodic GP, except using 24 walkers for 65,000 steps, excluding the first 15,000 as burn-in. This fit also resulted in a double-peaked posterior for $P_{\rm rot}$, with one peak at the upper bound of 15 days and the other at 14.1 days. The best-fit $P_{\rm dec}$ value is $114^{+28}_{-21}$ days. The corner plots of these fits are also shown in Figure B5.

Given these inconsistent results, we were not able to further constrain the stellar rotation period beyond the limits given by $v \sin i_\star$, and used the same uniform prior of 7.9–15 days when fitting the RVs for stellar activity. This inconclusive result is consistent with the findings of B. L. Martins et al. (2020), who labeled the stellar rotation signals as too low-amplitude to detect in this star's TESS light curves.

### *3.3.2. RV Analysis*

In addition to the magnetic cycle's effects, we also modeled the effects of starspots and the star's rotation when fitting the RVs with a GP model. We used a multidimensional quasiperiodic GP kernel, as implemented in PyORBIT, which follows the examples of V. Rajpaul et al. (2015) and O. Barragán et al. (2022). With this approach, we assume that the stellar-induced signals in the RV and the activity indicator can be described by the same latent GP and its time derivative. We denote the amplitudes of the RV as $V_r$ and $V_c$, and the amplitude of the Na II D activity index as $L_c$, as defined in V. Rajpaul et al. (2015) and following the notation of D. Nardiello et al. (2022). As noted in V. Rajpaul et al. (2015), these amplitudes can be related to physical quantities such as spot cover and contrast ($V_c$ and $L_c$) and stellar rotational and convective blueshift velocities ($V_r$), but are treated as free parameters (Table A2). The other hyperparameters for this kernel ($P_{\rm dec}$, $P_{\rm rot}$, and $w$), are the same as those used for the light-curve fitting. We set a uniform prior on $P_{\rm dec}$ of 50–4000 days—the lower bound comes from the TESS LCs, which show stable spot coverage over the span of at least 50 days, and the upper limit is set to approximately the time baseline of the RV observations, which allows for long-lived spots. As when fitting the light curves, we set bounds the on $P_{\rm rot}$ to 7.9–15 days as dictated by the $v \sin i_\star$ measurement, and set a prior of 0.5 ± 0.05 on $w$ following S. V. Jeffers & C. U. Keller (2009). We fit this GP model simultaneously with the Keplerian model described in Section 5.1. Using this multidimensional GP model, we detected a stellar rotation period of $9.42^{+0.14}_{-0.08}$ d. The best-fit GP model is plotted in Figure 2, with corner plots shown in Figure B6. The results are reported in Table A2.

Ultimately, we were not able to determine a single, coherent stellar rotation period for GJ 341. In addition to the low amplitude of the stellar rotation from the TESS OOT and ASAS-SN light curves, there is also a significant discrepancy between those $P_{\rm rot}$ values (approximately 13–15 days) and the $P_{\rm rot}$ value determined from the RVs and Na II D activity indicator (approximately 9 days). One thing to note is the nearly 5 yr offset between the final HARPS RV measurement and the first photometric measurement. This lack of coherence is consistent with the findings of L. Mignon et al. (2023)—a study of HARPS RVs and ASAS photometry of 177 M dwarfs with long-timescale variability. They also found that $P_{\rm rot}$ values determined from photometry and RVs taken noncontemporaneously to be highly incompatible.

## 4. Transit Fitting

### *4.1. TESS Transits*

We used batman (L. Kreidberg 2015) to model the TESS transits of GJ 341 b. We fit for the orbital period ($P$), transit time ($T_c$), planet radius divided by the stellar radius ($R_p/R_\star$), impact parameter (b), stellar density ($\rho_\star$), stellar quadratic limb-darkening coefficients ($u_1$, $u_2$), and jitter, assuming a circular orbit for the planet. We set uninformative, uniform priors on all but P and $T_c$ (see Table A1). For those two parameters, our uniform priors were informed by the values reported in the TOI catalog,[16] but with upper and lower bounds well beyond 1$\sigma$ from the reported values (approximately 250$\sigma$ from the reported $P$ values and 30$\sigma$ from the reported $T_c$ value). We used emcee (D. Foreman-Mackey et al. 2013), running the sampler for 100,000 steps, excluding the first 32,000 as burn-in, and using a thinning factor of 100. All posteriors are plotted in Figure B7.

[16] https://tev.mit.edu/data/

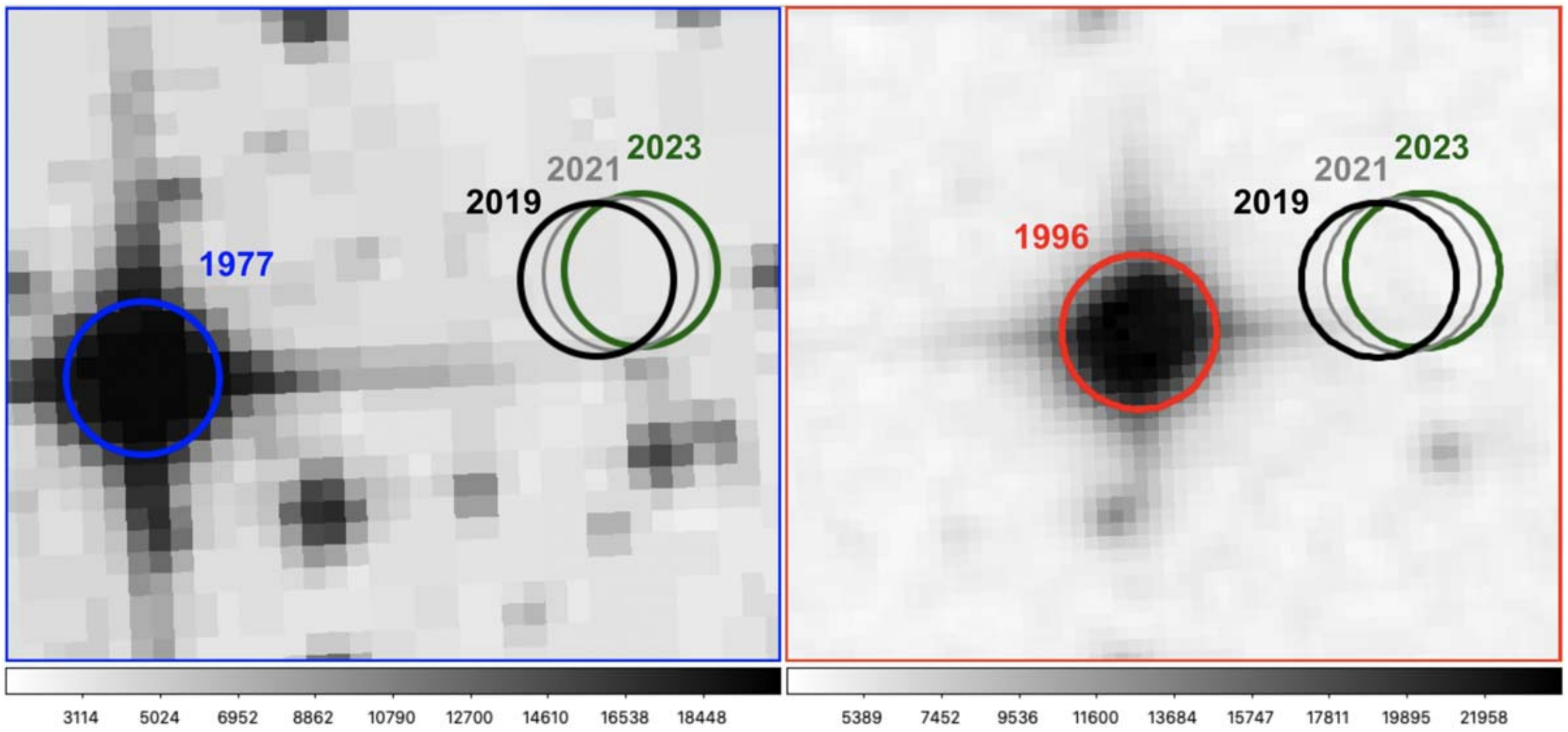

**Figure 6.** POSS2/UKSTU Blue image (3400–5900 Å, left) and POSS2/UKSTU Red image (5900–7150 Å, right), both centered on GJ 341. The blue (left) and red (right) circles indicate the location of the star at the time of those images, in 1977 and 1996, respectively. The black circles indicates the location of the star at the start of its first TESS observation, and the gray circles indicate its location at the time of its last TESS observation (2019 February and 2021 April, respectively). The green circle indicates the location of the star at the time of the JWST observations, in 2023 March. In both images, we see no indication of a background source that would have been hidden behind GJ 341, and therefore uncorrected, at the time of the TESS or JWST observations. Although some of the surrounding stars fall onto the same 21″ × 21″ TESS pixel as GJ 341, their flux was corrected for in the report of the TESS photometric measurements of GJ 341. The annotated annuli are 6" in radius, and their locations were determined using the Gaia DR3 astrometry and proper motion values (Gaia Collaboration et al. 2023), and extrapolating them from the Gaia epoch of 2016 to the time of their observations. The top of this image is north and the left of this image is east.

We fit this model to 11 TESS transits simultaneously, each with 6 hr of observations centered on the transit. We used 20 hr of baseline, with 3 hr for the transit cut out, to calculate the slope of the baseline for each transit, then subtracted this slope from each transit before fitting. These results are plotted in the left-side panel of Figure 3 and listed in Table 2. Our fit period and transit time are consistent with those noted on exoMAST.[17]

### 4.2. Possible Blending

TESS has large pixels, 21″ × 21″, and GJ 341 is in a relatively crowded field (G. R. Ricker et al. 2015; J. C. Smith et al. 2012; M. C. Stumpe et al. 2012). This leads to the possibility that the transit is blended with uncorrected flux from a nearby star. We found, however, using POSS2/UKSTU images in combination with astrometry and proper motion measurements from Gaia, that there is no measurable flux in the background of GJ 341 at the time of the TESS observations nor at the time of the JWST observations (between 2019 February and 2021 April nor in 2023 March; see Figure 6). Therefore, the surrounding sources are known and corrected for in the TESS light curves and there is no blending in the JWST observations.

### 4.3. JWST Transits

Like with the TESS light curves, we used `batman` (L. Kreidberg 2015) to fit the JWST transit white light curves of GJ 341 b. However, we also needed to fit an additional systematics model to account for the long-timescale trend in the data. Each of our three reductions of the JWST data fitted this trend with an exponential ramp multiplied by a linear polynomial. For our `tswift` reduction, we fitted the white light from each NIRCam amplifier with its own systematics model. Each pipeline made a different choice about whether to clip integrations from the start of the time series, with `Eureka!` clipping 300 while `Tiberius` and `tswift` clipped no integrations. A full description of the fitting choices made in our JWST analysis are given in J. Kirk et al. (2024). We fitted the three JWST transits simultaneously and fixed the period to the value derived from the TESS transits fit in Section 4.1. We fitted for $T_0$, $i$, $a/R_*$, and $R_p/R_*$, and the parameters of the systematics model. For the limb-darkening calculations we assumed solar metallicity for the star ([Fe/H] = 0). Using GJ 341's measured metallicity of [Fe/H] = −0.16 results in limb-darkening coefficient (LDC) variations of $10^{-4}$ when compared to solar metallicity (orders of magnitude below typical LDC measurement uncertainties), so we proceed with solar metallicity as a fair approximation for the LDC calculations. We adopted values of $T_{\rm eff}$ and $\log g$ as derived in Section 3.1. We used a quadratic limb-darkening law, with coefficient values fixed to those calculated using ExoTiC-LD (D. Grant & H. R. Wakeford 2022) and a custom instrument throughput obtained by dividing the observed stellar spectrum by the `PHOENIX` model closest to the derived parameters for the star (T. O. Husser et al. 2013). The best-fitting system parameters from each pipeline are all in agreement with one another as well as with the TESS results (with the exception of $R_p/R_\star$ as fit from the `Tiberius` pipeline), and all are given in Table 2.

## 5. RV Analysis for Planetary Signal

### 5.1. Keplerian

We added a Keplerian to the RV model to fit for GJ 341 b simultaneously with the GP model for stellar activity (as described in Section 3.3.2). We set tight priors on the period and transit time of the planet, using the results from the TESS LC fit. We used `emcee` to fit the joint stellar and planet model, running the sampler for 850,000 steps, with 300,000 steps burn-in and a thinning factor of 100. The corner plots for the GP and keplerian models are shown separately in Figures B6 and B9, and all parameters are presented in Table A2. We find a semiamplitude for the planet consistent with 0 m s$^{-1}$, meaning that we cannot measure a mass for the planet using the existing HARPS RV data set. This result falls within expectations. A planet composed of pure iron with a radius of 0.88 $R_\oplus$, like GJ 341 b, will have a mass of less than 1.3018 $M_\oplus$, which would induce a semiamplitude of less than 0.69 m s$^{-1}$ in the star's RVs (L. Zeng et al. 2019). This is beyond the RV precision limit set by photon noise for this star as observed by HARPS. The phase-folded RV curve of this result is plotted in the right panel of Figure 3.

[17] https://exo.mast.stsci.edu/

### 5.2. Upper-limit Mass Measurement

Given the nondetection of the planetary signal, we set out to determine upper limits for the planetary mass. We did this by adding Keplerian models for planets with masses between 0.5 and 8.0 $M_\oplus$ (with a step size of 0.5 $M_\oplus$ between 0.5 and 2.0 $M_\oplus$ and 4.5–8.0 $M_\oplus$, and 0.1 $M_\oplus$ between 2.0 and 4.5 $M_\oplus$), to the observed HARPS RVs, assigning the amplitudes of the signals, while fixing all other parameters of the Keplerian to those obtained from the TESS LC fit. We then varied the calculated RV at each time of observation by randomly assigning an uncertainty value from any of the observed HARPS RV points. Then we drew from a Gaussian distribution centered on the data value plus the Keplerian model, with a standard deviation equal to that assigned uncertainty. This drawn value was taken as the RV value at that time stamp, and the randomly assigned uncertainty was taken as the measurement uncertainty.

From there, we used PyORBIT to fit a semiperiodic GP kernel and a Keplerian model to this generated data set. We fixed all of the parameters and hyperparameters to our previously fit values (see Table A2), except for the Keplerian semiamplitude, the amplitude of the GP, the jitter and the offset of the data. Our aim was to see if we would fit a semiamplitude for the Keplerian that would constitute a planet detection, while allowing for the possibility that the planet signal may be absorbed into either the GP amplitude or the jitter term. We also allowed the offset to vary in order to fit our new randomized data set. We ran this trial 500 times per injected mass, and evaluated how many times we fit a mass consistent with a 3$\sigma$ detection of a planet. This value, divided by the total number of trials, is represented as our Recovery Rate in Figure 7. From Figure 7 we infer a 4.0 $M_\oplus$ 3$\sigma$ upper limit and a 2.9 $M_\oplus$ 1$\sigma$ upper limit for the mass of GJ 341 b. This is well within the planetary mass expectations, as highlighted in Section 5.1.

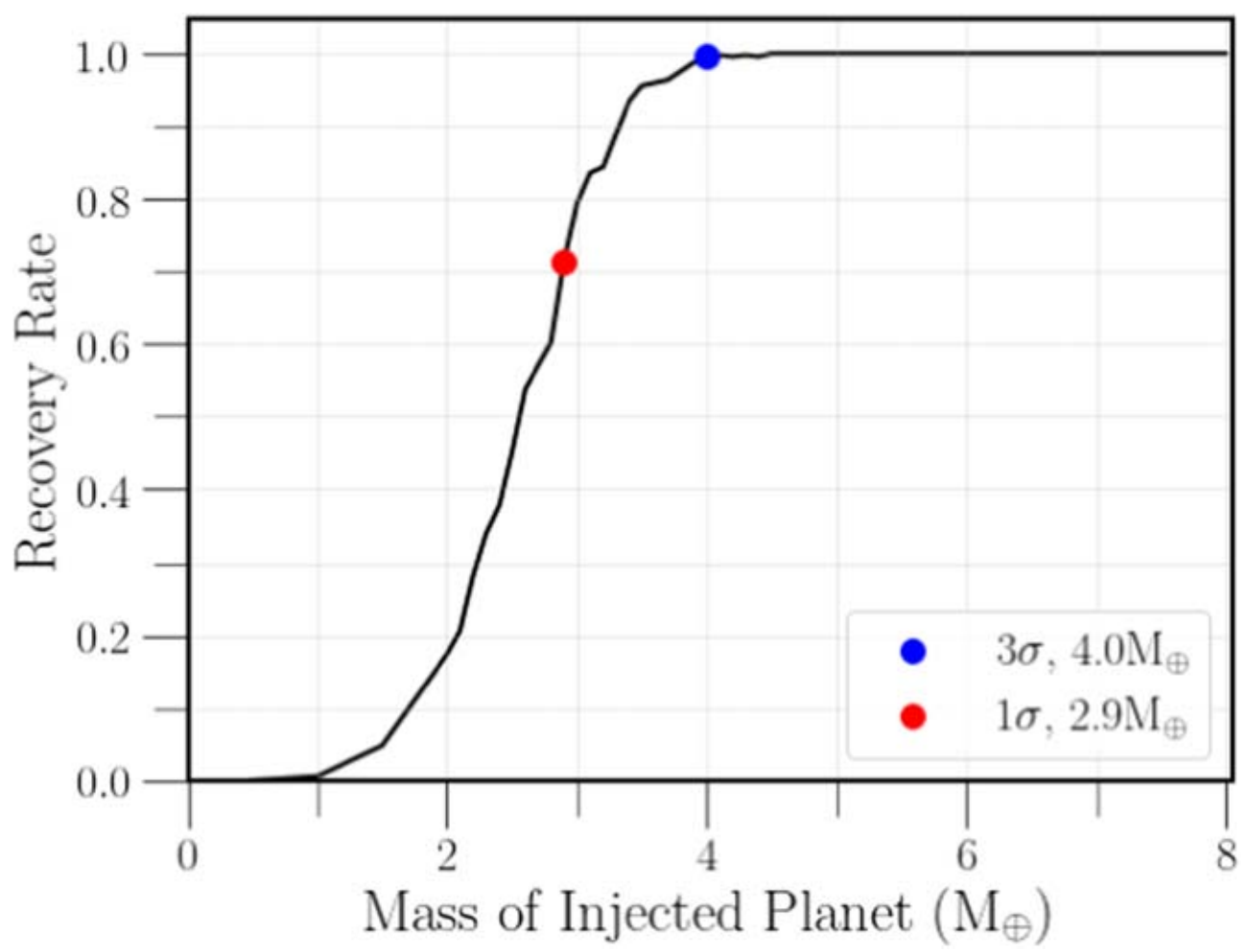

**Figure 7.** The proportion of planets recovered when fitting for Keplerian signals injected into the HARPS RVs. We detect injected planets with $M_p$ = 2.9 $M_\oplus$ at 7.577 days for 71.4% of the trials, and therefore take this mass as our 1$\sigma$ upper-limit mass measurement (marked with a red point). We recover injected planets with $M_p$ = 4.0 $M_\oplus$ at 7.577 days for 99.8% of our trials, and therefore take this mass as our 3$\sigma$ upper-limit mass measurement (marked with a blue point).

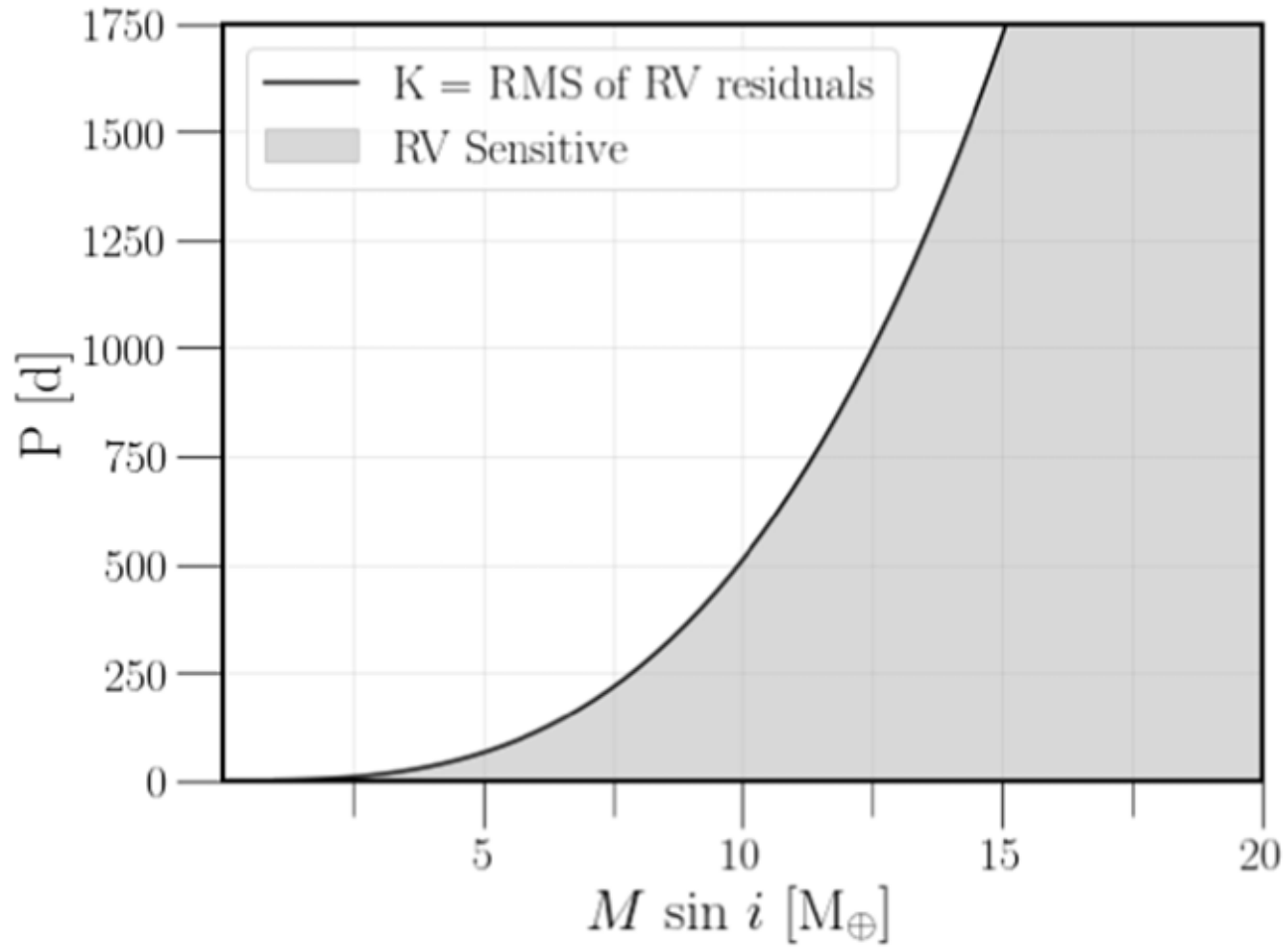

**Figure 8.** Period in days plotted vs. $M$ sin $i$ for companions that would impart a semiamplitude equal to that of the rms of the residuals after we remove our best-fit GP model from the RV data set (rms = 1.31 m s$^{-1}$, black line). Based on the RV data set and our best-fit model, we rule out the presence of a companion with an $M$ sin $i$ greater than 15.1 $M_\oplus$ out to a period of 1750 days (one-half of the time baseline of the RV data set).

### 5.3. Additional Companions

We also used the existing HARPS RVs to place $M$ sin $i$ limits for additional stellar or substellar companions in the system at orbital periods smaller than $P$ = 1750 days (half of the RV time baseline). We calculated the $M$ sin $i$ and period of companions that lie just at the limit of RV detectability, as dictated by the rms of the RV residuals (1.31 m s$^{-1}$), and represent this result in Figure 8. The gray area on this plot indicates the $M$ sin $i$ and period space that we are sensitive to, as companions in this space would have imparted an RV semiamplitude above the residuals rms. We rule out the presence of a companion with an $M$ sin $i$ greater than 15.1 $M_\oplus$ out to a period of 1750 days (one-half of the time baseline of the RV data set).

## 6. Results and Discussion

In this work we perform a detailed analysis of available photometric, spectroscopic, and astrometric data for the GJ 341 system to help establish the planetary nature of GJ 341 b, and provide improved system parameters to help characterize its atmosphere with recently obtained JWST observations. We measure a new radius for the host star of $R_\star = 0.5066^{+0.0169}_{-0.0172} R_\odot$, which combined with the analysis of the TESS light curves, yields an updated planet radius of $R_p = 0.88 \pm 0.05 R_\oplus$. The transit parameters fit from the TESS observations agree well with those fit from three different reductions of the JWST white light curves, with the exception of $R_p/R_\star$ as fit from the Tiberius pipeline light curve. We establish that the star is not in a binary, and any additional companions with periods shorter than 1750 days must be less massive than $m \sin i = 15.1\ M_\oplus$. We establish that GJ 341 was not blended with any background sources during either the TESS or the JWST observations. We also present evidence of the star's magnetic cycle.

We were not able to detect the planetary signal in the existing HARPS RV observations, but place strong upper limits of less than 4.0 $M_\oplus$ ($3\sigma$), and less than 2.9 $M_\oplus$ ($1\sigma$) for its mass. Even a pure iron planet of that radius would have a mass of less than 1.3018 $M_\oplus$ (L. Zeng et al. 2019). Such a planet would induce a semiamplitude of less than 0.69 m s$^{-1}$ in the star's RVs. Although this is just beyond the RV precision limit set by photon noise for this star as observed by HARPS, this planet will likely be detectable with a higher-precision instrument like ESPRESSO (F. Pepe et al. 2014), which has achieved $\leqslant$0.7 m s$^{-1}$ precision RV measurements for nearby early M dwarfs that are comparable to GJ 341 (J. Lillo-Box et al. 2021; B. Lavie et al. 2023; A. Castro-González et al. 2023).

GJ 341 was included in previous TESS candidate validation studies—namely S. Giacalone et al. (2020) and S. Palatnick et al. (2021)—neither of which favored the candidate to be a true planet. We do not, however, consider these results to be in conflict with our findings. Using the TESS observations and their vetting code, TRICERATOPS, S. Giacalone et al. (2020) did not find these transits likely to belong to a planet nor likely to be a nearby false positive. In that paper, they considered the nearby TESS Input Catalog (TIC) stars, their positions, TESS magnitudes, and stellar properties in order to determine the likelihood that the TESS transit observations originated from a planet around the target star and that the transit was not being diluted by an unresolved background star. An updated version of TRICERATOPS, which incorporates the JWST transit light curves and uses the K-band to approximate NIRCam bandpass photometry, finds an updated false-positive probability of $1.7 \times 10^{-9}$, which is well below the typical planet validation cutoff of $1 \times 10^{-2}$, and an updated nearby false-positive probability of zero (J. Gomez Barrientos and S. Giacalone, private communication).

Using both the TESS observations and the HARPS RVs we use in this work, S. Palatnick et al. (2021) assigned this a false-alarm probability of 92.6%. Our work is consistent with this finding in that we were not able to detect GJ 341 b in the available HARPS measurements. As a comment on the high false-alarm probability, however, we do note that S. Palatnick et al. (2021) detrended the RVs of the stars in their sample using a flat line, linear trend or quadratic trend, as was appropriate. As we note in Section 3.2, the long-term trend of these RVs is better fit by a higher-order polynomial, so the approach to modeling long-term trends that S. Palatnick et al. (2021) used, although generally applicable, was not well suited for this particular target.

Despite the lack of RV detection, we confirm that this planet transits the correct star via its JWST transit observations. The three transits observed by JWST matched the transit depth and timing as they were predicted from the TESS observations, dispelling concerns of contamination or false positives.

## Acknowledgments

This material is based upon work supported by the National Science Foundation Graduate Research Fellowship under Grant No. DGE1745303 and by NASA under award No. 80GSFC21M0002. This publication is funded in part by the Alfred P. Sloan Foundation under grant G202114194.

This work has made use of data from the European Space Agency (ESA) mission Gaia (https://www.cosmos.esa.int/gaia), processed by the Gaia Data Processing and Analysis Consortium (DPAC, https://www.cosmos.esa.int/web/gaia/dpac/consortium). Funding for the DPAC has been provided by national institutions, in particular the institutions participating in the Gaia Multilateral Agreement. This work is based on observations from the HARPS spectrograph collected at the European Southern Observatory under ESO programs: 081.C-0802(B), 072.C-0488(E), 183.C-0437(A), 183.C-0972(A), 090.C-0421(A), 086.C-0284(A), 089.C-0732(A), 087.C-0831(A), 082.C-0390(A).

This paper includes data collected by the TESS mission, which are publicly available from the Mikulski Archive for Space Telescopes (MAST). Funding for the TESS mission is provided by NASA's Science Mission Directorate.

The JWST observations are associated with program #1981. Support for program #1981 was provided by NASA through a grant from the Space Telescope Science Institute, which is operated by the Association of Universities for Research in Astronomy, Inc., under NASA contract NAS 5-03127.

We thank Jonathan D. Gomez Barrientos and Steven Giacalone for their support in using the TRICERATOPS package to calculate an update false-positive probability and nearby false-positive probability for this target. We thank Emily Pass for helpful and productive discussion about determining the rotation periods of M dwarfs.

The Digitized Sky Surveys were produced at the Space Telescope Science Institute under U.S. Government grant NAG W-2166. The images of these surveys are based on photographic data obtained using the Oschin Schmidt Telescope on Palomar Mountain and the UK Schmidt Telescope. The plates were processed into the present compressed digital form with the permission of these institutions. The Second Palomar Observatory Sky Survey (POSS-II) was made by the California Institute of Technology with funds from the National Science Foundation, the National Geographic Society, the Sloan Foundation, the Samuel Oschin Foundation, and the Eastman Kodak Corporation. The Oschin Schmidt Telescope is operated by the California Institute of Technology and Palomar Observatory. The UK Schmidt Telescope was operated by the Royal Observatory Edinburgh, with funding from the UK Science and Engineering Research Council (later the UK Particle Physics and Astronomy Research Council), until 1988 June, and thereafter by the Anglo-Australian Observatory. The blue plates of the southern Sky Atlas and its Equatorial Extension (together known as the SERC-J), as well as the Equatorial Red (ER), and the Second Epoch [red] Survey (SES) were all taken with the UK Schmidt. Supplemental funding for sky-survey work at the STScI is provided by the European Southern Observatory.

## Appendix A
## Tables

In this appendix, we provide additional tables which detail the analysis and results from the text. Table A1 presents the priors and posteriors relevant to our fits of the TESS transit light curves, and Table A2 presents the priors and posteriors relevant to our fits of the HARPS RV and Na II D measurements.

**Table A1**
Transit Model Priors and Posteriors

| Parameter | Prior | Posterior |
|---|---|---|
| Sampled Parameters | | |
| $P$ (d) | $\mathcal{U}$: (7.0, 8.0) | $7.576860^{+0.000034}_{-0.000020}$ |
| $R_p$ ($R_\star$) | $\mathcal{U}$: (0.00001, 0.5) | 0.0160 ± 0.0008 |
| $T_c$ (BJD-2.45e6) | $\mathcal{U}$: (9301.650, 9301.850) | 9301.771 ± 0.002 |
| b | $\mathcal{U}$: (0.00,0.90) | $0.33^{+0.29}_{-0.23}$ |
| $\omega$ (deg) | Fixed | 90.0 |
| $e$ | Fixed | 0.0 |
| Stellar Density ($\rho_\odot$) | $\mathcal{U}$: (0.00, 5.00) | $3.46^{+0.94}_{-1.40}$ |
| $u_1$ | $\mathcal{U}$: (0.00, 1.00) | $0.39^{+0.38}_{-0.29}$ |
| $u_2$ | $\mathcal{U}$: (−1.00, 1.00) | $0.07^{0.38}_{-0.29}$ |
| Jitter | $\mathcal{U}$: (0.000004, 0.043410) | 0.00023 ± 0.00001 |
| Derived Parameters | | |
| a (au) | ... | $0.058^{+0.005}_{-0.009}$ |
| $i$ (deg) | ... | $89.24^{+0.53}_{-0.96}$ |
| $R_p$ ($R_\oplus$) | ... | 0.88 ± − 0.05 |
| Transit Duration (hr) | ... | $2.23^{+0.12}_{-0.08}$ |

**Table A2**
Priors and Posteriors for the Models Fit to the HARPS RVs and the Na II D Measurements

| Parameter | Prior | Posterior |
|---|---|---|
| Gaussian Process Parameters | | |
| $V_c$ ( m s$^{-1}$) | $\mathcal{U}$: (0, 20) | $2.2^{+0.6}_{-0.5}$ |
| $V_r$ ( m s$^{-1}$) | $\mathcal{U}$: (−20, 20) | $0.3^{+1.0}_{-1.1}$ |
| $L_c$ ( m s$^{-1}$) | $\mathcal{U}$: (−500, 500) | $0.008^{+0.002}_{-0.001}$ |
| $P_{\rm dec}$ (d) | $\mathcal{U}$: (50.0, 4000.0) | $352^{+162}_{-176}$ |
| $P_{\rm rot}$ (d) | $\mathcal{U}$: (7.9, 15.0) | $9.42^{+0.14}_{-0.08}$ |
| $w$ | $\mathcal{N}$: 0.5 ± 0.05 | 0.52 ± 0.05 |
| Keplerian Model Parameters | | |
| K ( m s$^{-1}$) | $\mathcal{L}_2\mathcal{U}$: (−9.97, 10.97) | $0.016^{+0.11}_{-0.014}$ |
| Tc (BJD-2.45e6) | $\mathcal{N}$: 9301.771 ± 0.002 | 9301.771 ± 0.002 |
| $P$ (d) | $\mathcal{N}$: 7.576860 ± 0.000034 | 7.576860 ± 0.000034 |
| $\omega$ (deg) | Fixed | 90 |
| e | Fixed | 0 |
| Data Offset and Jitter | | |
| RV offset ( m s$^{-1}$) | $\mathcal{U}$: (−10000,10000) | $-0.43^{+0.78}_{-0.79}$ |
| Na II D offset | $\mathcal{U}$: (−10000,10000) | 0.290 ± 0.003 |
| RV jitter ( m s$^{-1}$) | $\mathcal{U}$: (0.007, 200.5) | $1.64^{+0.31}_{-0.29}$ |
| Na II D jitter | $\mathcal{U}$: (0.000, 0.470) | 0.003 ± 0.001 |

**Note.** These are the parameters and hyperparameters for the quasiperiodic multidimensional GP, fit to the HARPS RV and Na II D measements, as well as the Keplerian model fit simultaneously to the HARPS RVs. Ultimately, these fits result in a nondetection of the 7.577 day period planet in this RV data set.

## Appendix B
## Figures

In this appendix, we include additional figures presenting secondary data and analysis products relevant to the main text. We plot the ASAS-SN light curve for GJ 341, as discussed in Section 2.2, and our best-fit GP model in Figure B1. We plot the HARPS activity indicators versus the HARPS RVs to show their correlations, or lack thereof, as discussed in Section 3.2 in Figure B2. In Figure B3, as referenced in Section 2.4, we present the additional reductions of the JWST light curves, phase-folded to the period of the 7.577 day planet. In Figure B4, we present the sensitivity curves from the three studies that have searched for companions to GJ 341. We present the corner plots resulting from our MCMC fits of a quasiperiodic kernel fit to the available OOT light curves (Figure B5), of a multidimensional quasiperiodic kernel fit to the HARPS RVs and Na II D measurements (Figure B6), and of our transit model fit to the TESS in-transit light curves (Figure B7). We show periodograms for the HARPS RVs and available activity indicators in Figure B8. Finally, we show the corner plots from our MCMC fit of the Keplerian model to the HARPS RVs in Figure B9.

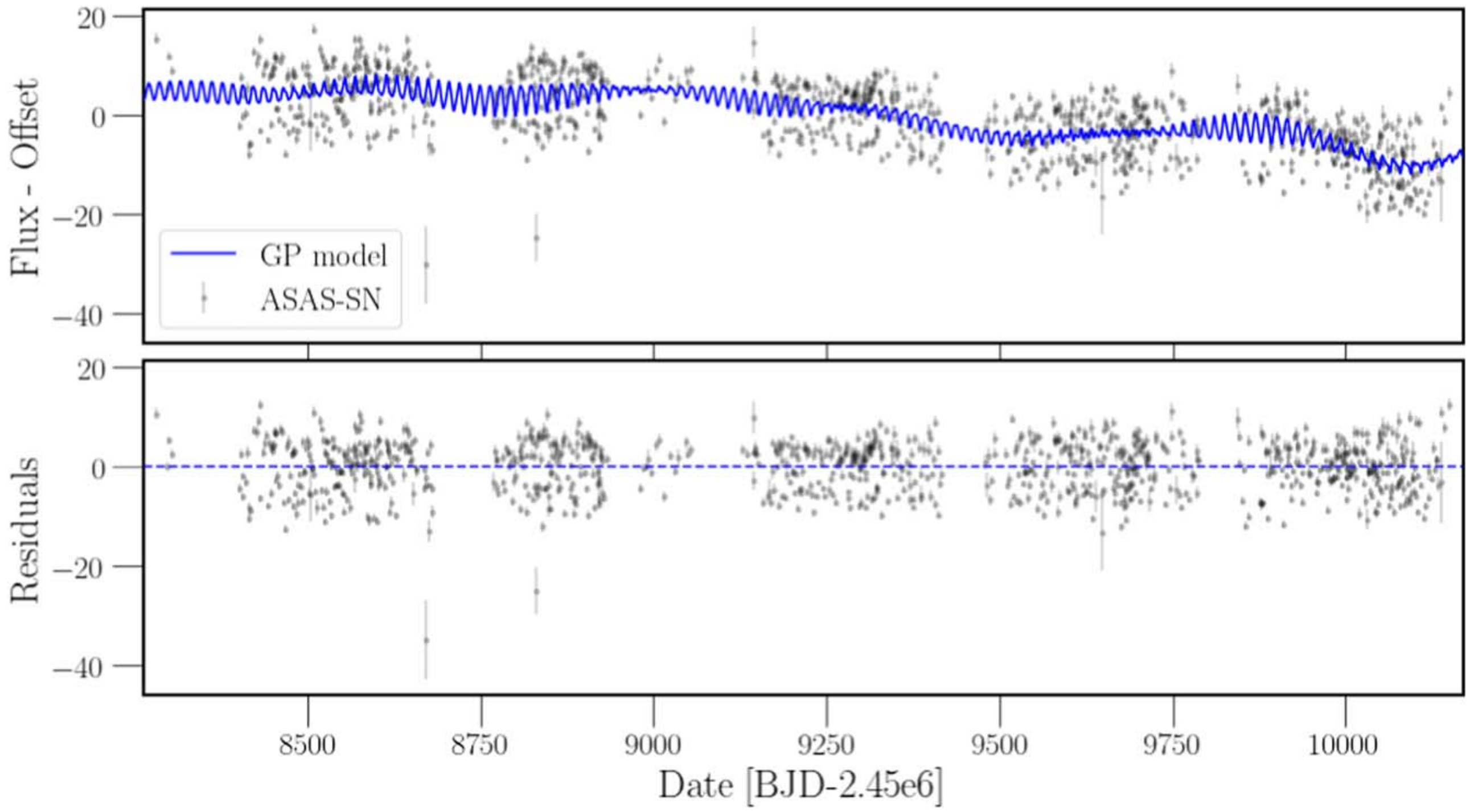

**Figure B1.** ASAS-SN light curve for GJ 341, which was observed from 2018 June to 2023 July. When fitting our GP model (blue), we binned the data down to a cadence of one point per 12 hr, as represented in this figure. The residuals of the best-fit GP model is plotted in the lower panel.

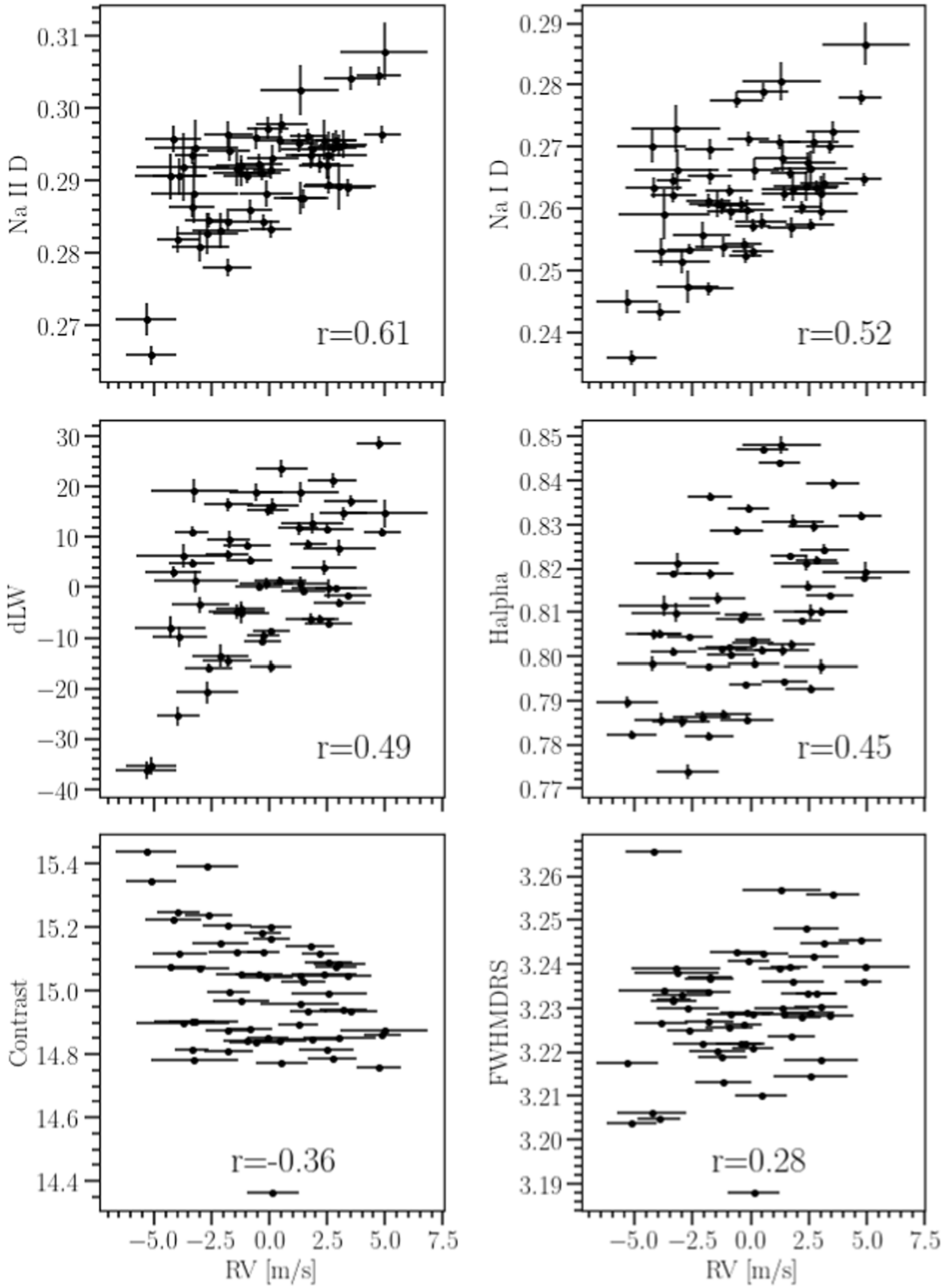

**Figure B2.** Various HARPS stellar activity indicators vs. GJ 341's RVs. The r-values are Pearson correlation coefficients. All of this star's activity indicators are moderately or significantly correlated with its RVs.

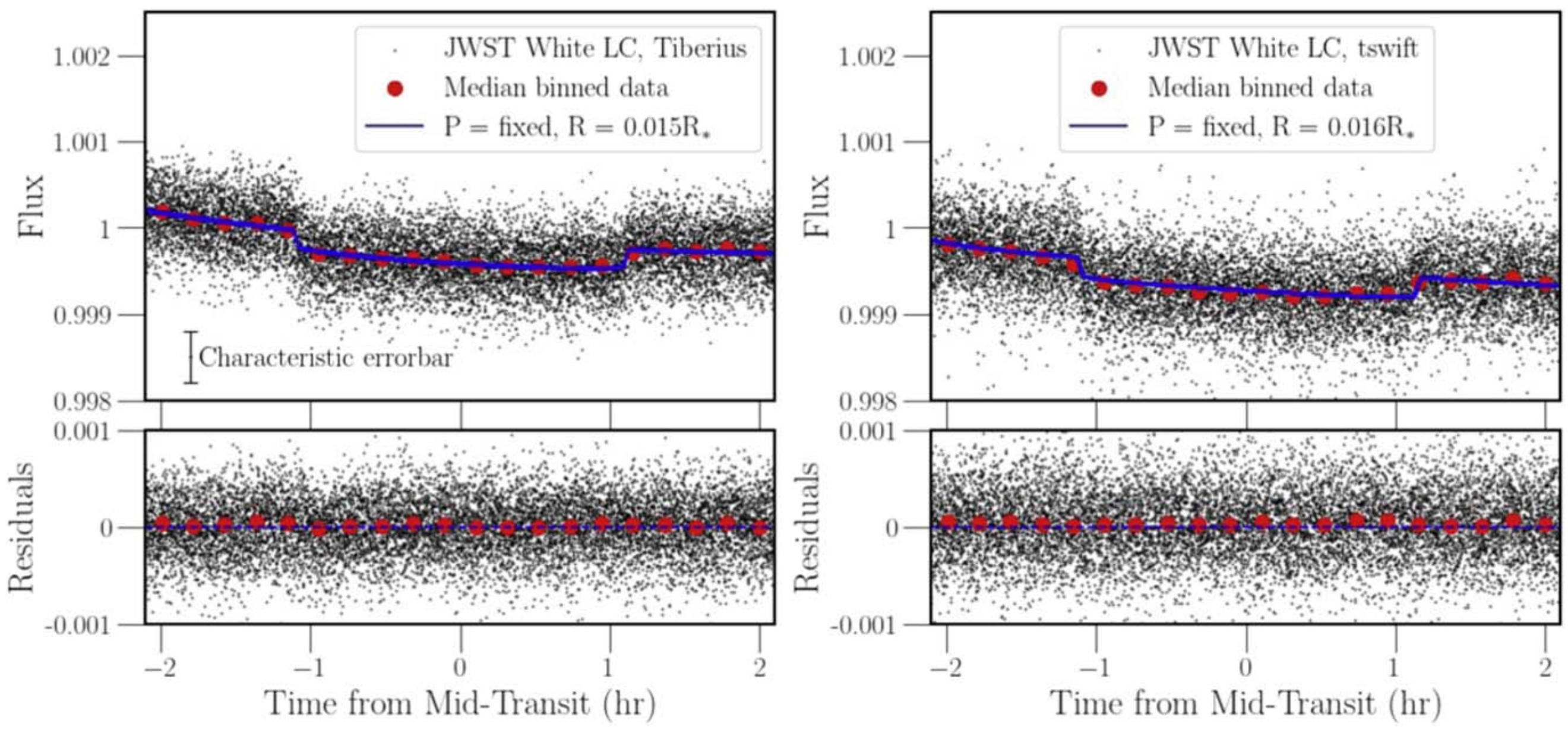

**Figure B3.** Phase-folded fit of the JWST transits, as produced by the `Tiberius` (left) and `Tswift` (right) reduction pipelines. Both panels show the phase-folded JWST light curve in black, including a characteristic error bar, the circular orbit transit model in blue, and the median data value within a bin of 0.2 hr in red. The upper portion shows the model plotted over the data and the lower portion shows the residuals of the data minus the model.

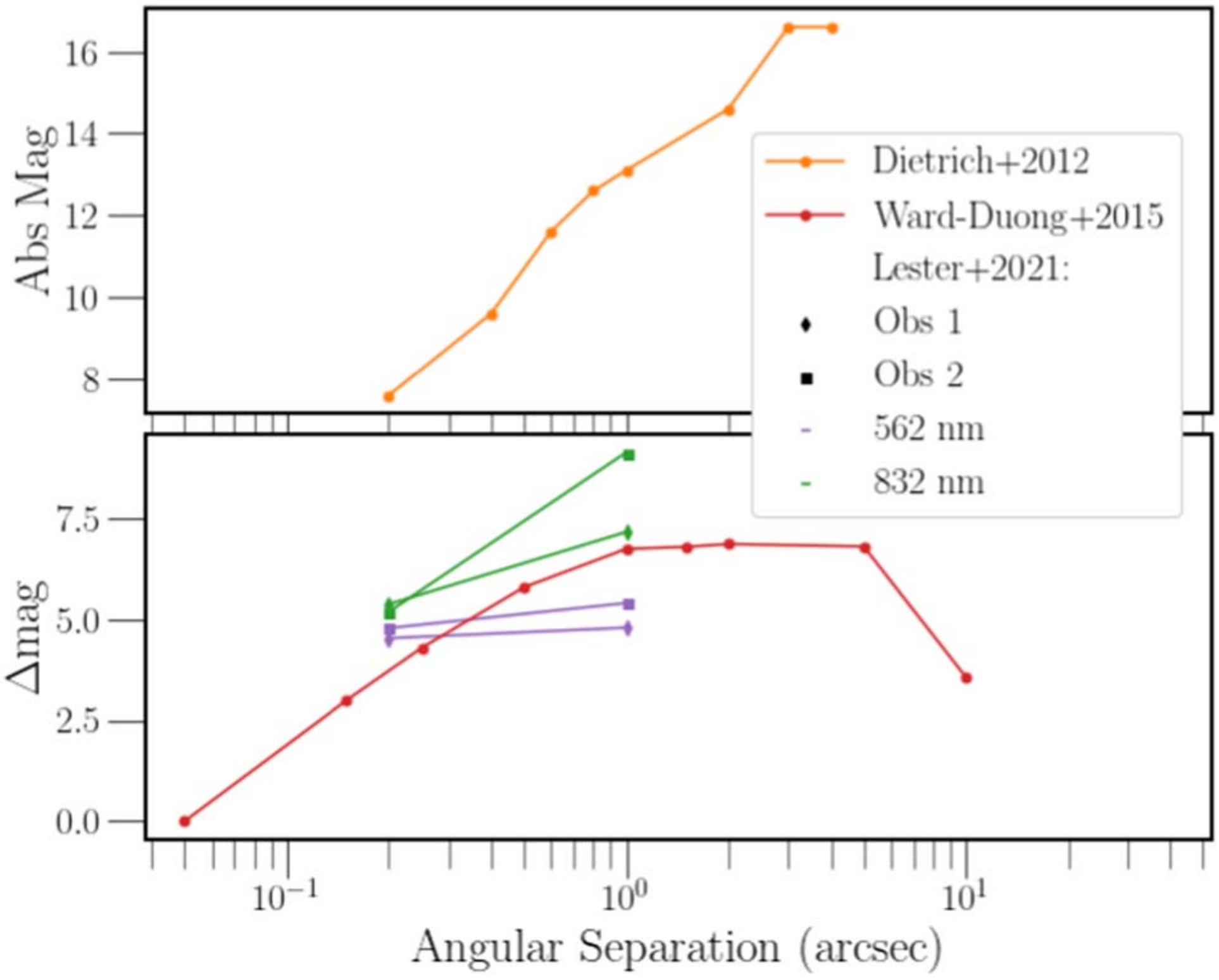

**Figure B4.** Sensitivity curves from S. B. Dieterich et al. (2012), K. Ward-Duong et al. (2015), and K. V. Lester et al. (2021). S. B. Dieterich et al. (2012) used high-resolution imaging with Hubble Space Telescope/NICMOS to place limits on companions around M dwarfs, and quoted their sensitivity limits in terms of the absolute F180M ($\lambda_{\rm ref} = 19004.28$ Å) magnitude (orange line, upper panel). K. Ward-Duong et al. (2015) used the NaCo/S13 instrument on Very Large Telecscope (NB1.64 filter, $\lambda_c = 1.644\mu$m) adaptive optics imaging to place constraints on companions around GJ 341, quoted in magnitude contrast (red line, lower panel). K. V. Lester et al. (2021) used the ‘Alopeke and Zorro speckle cameras at Gemini South to perform speckle imaging on GJ 341. They took two sets of observations and report sensitivities in terms of magnitude contrast at two different wavelengths for each observation (diamonds indicate their first observation, squares indicate their second observation, purple indicates their sensitivity at 562 nm and green indicates their sensitivity at 832 nm, lower panel).

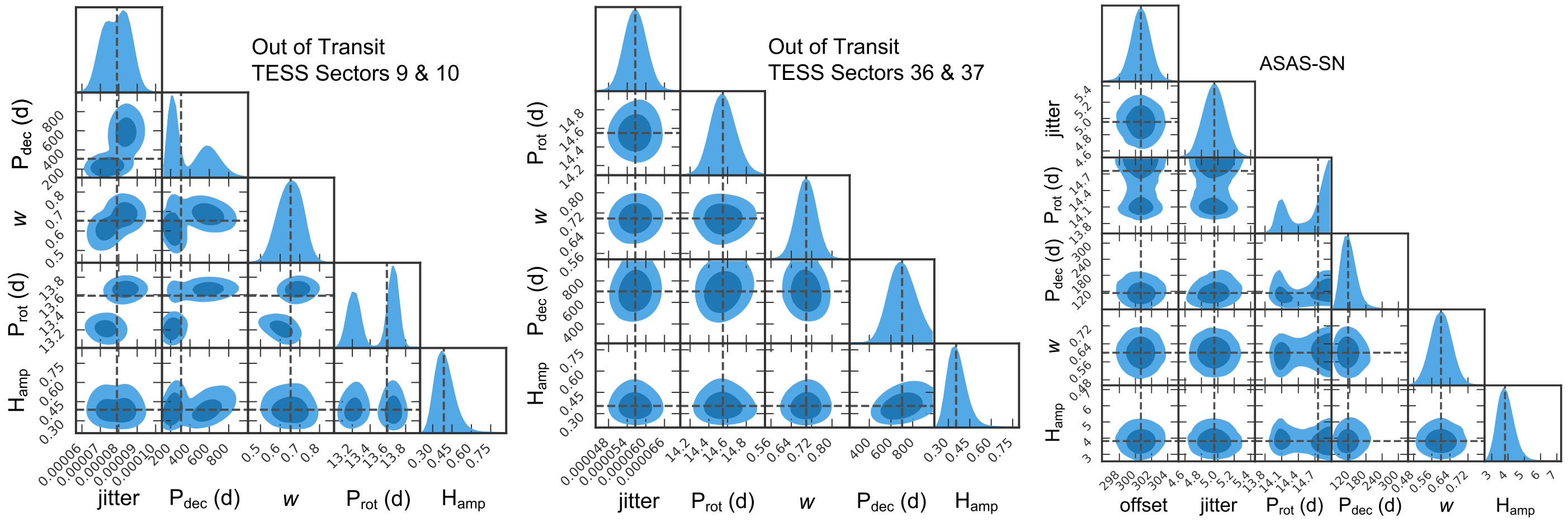

**Figure B5.** Corner plots from the fit of a quasiperiodic GP kernel to out of transit TESS and ASAS-SN light curves. We fit TESS sectors 9 and 10 (left) separately from sectors 36 and 37 (center). Although all three best-fit $P_{\rm rot}$ are in agreement with the $P_{\rm rot}$ derived from the v sin *i*published in S. Hojjatpanah et al. (2019), the inconsistency between the data sets means that we cannot place any further constraints on $P_{\rm rot}$ when fitting the RVs based on these LC fits.

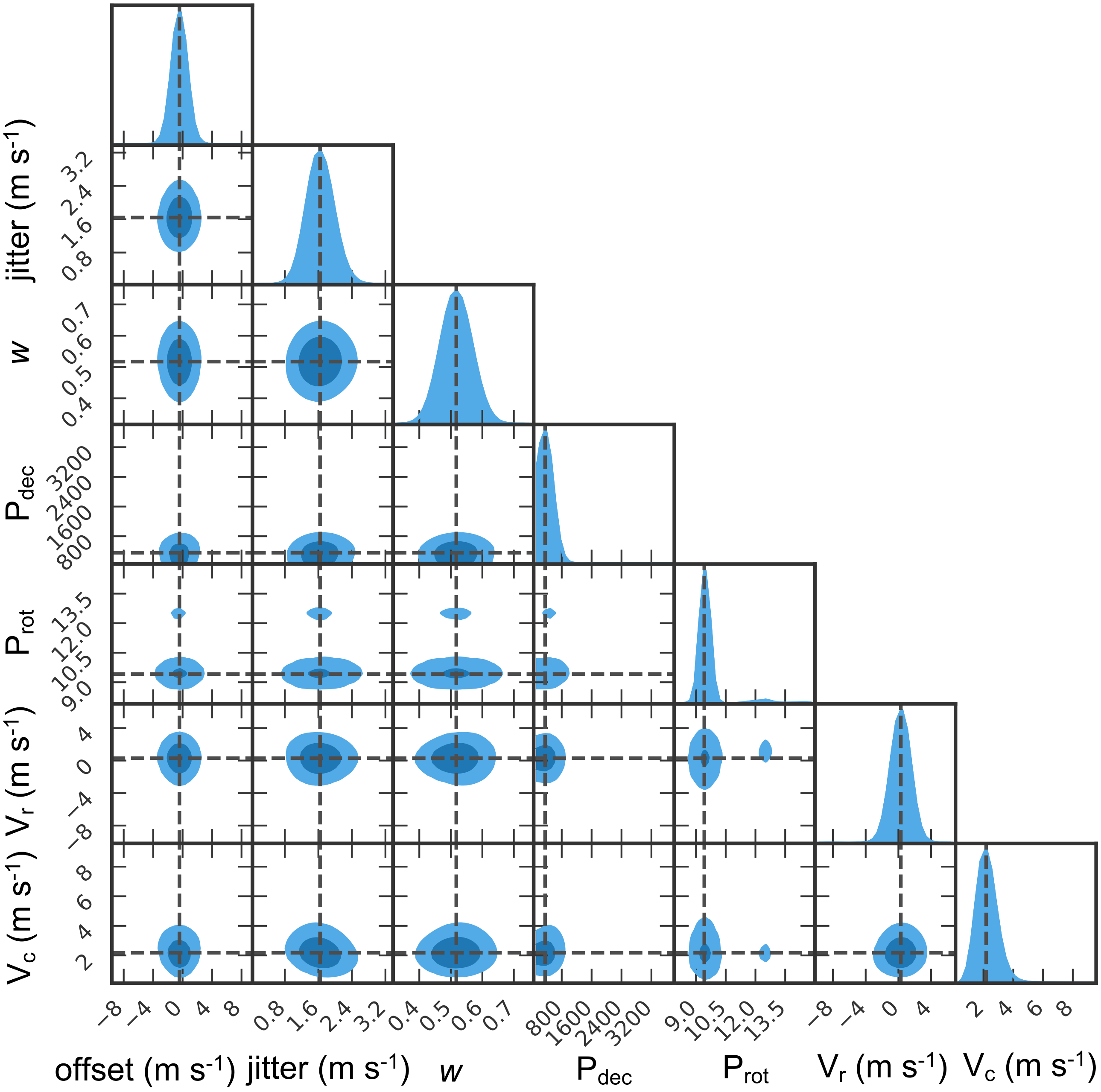

**Figure B6.** Corner plots for RV data fit with a multidimensional, quasiperiodic GP. These parameters were fit simultaneously with the Keplerian model, but corner plots are presented separately for readability.

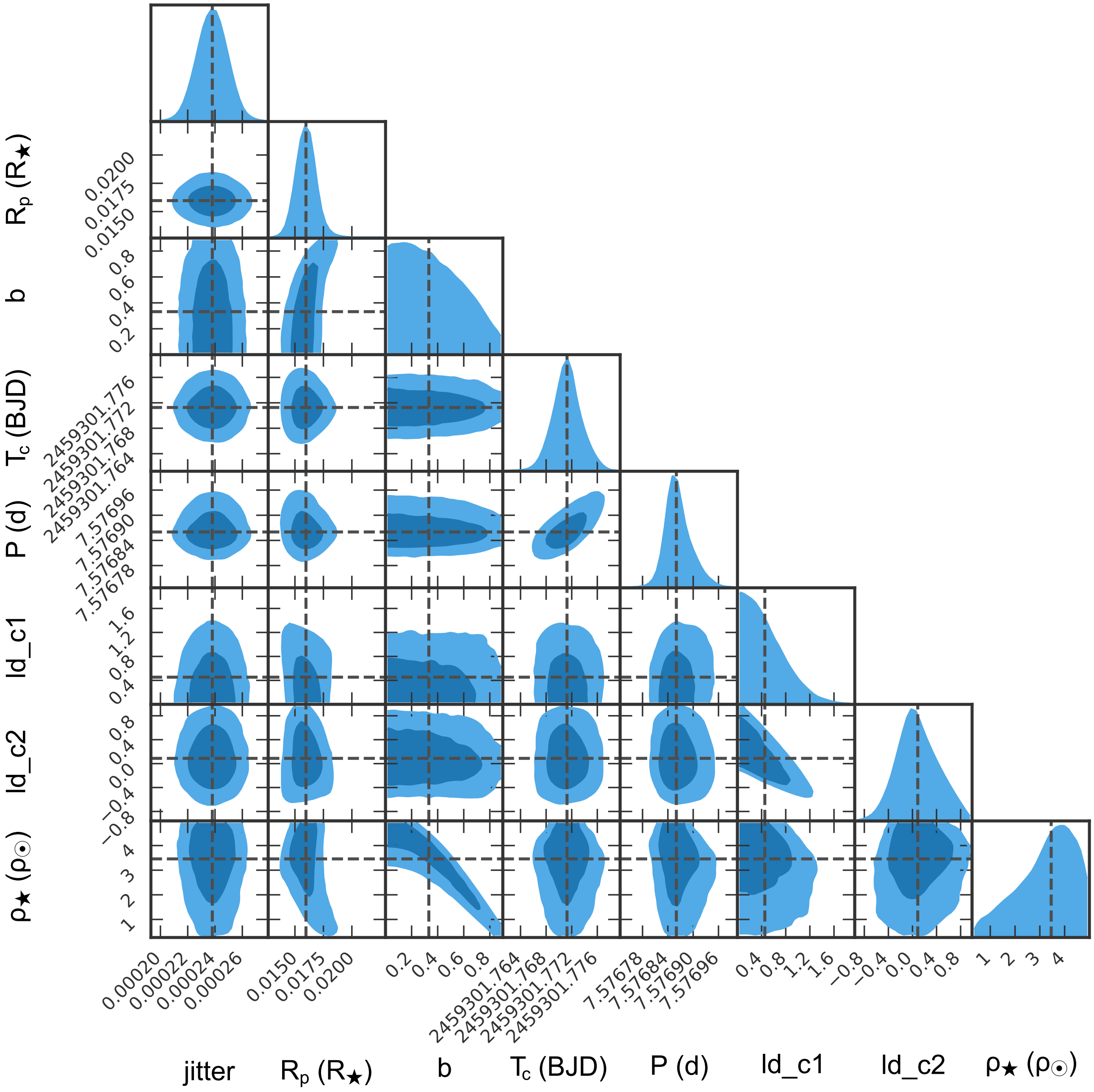

**Figure B7.** Corner plots for in-transit light curve fit with a transit and limb-darkening model.

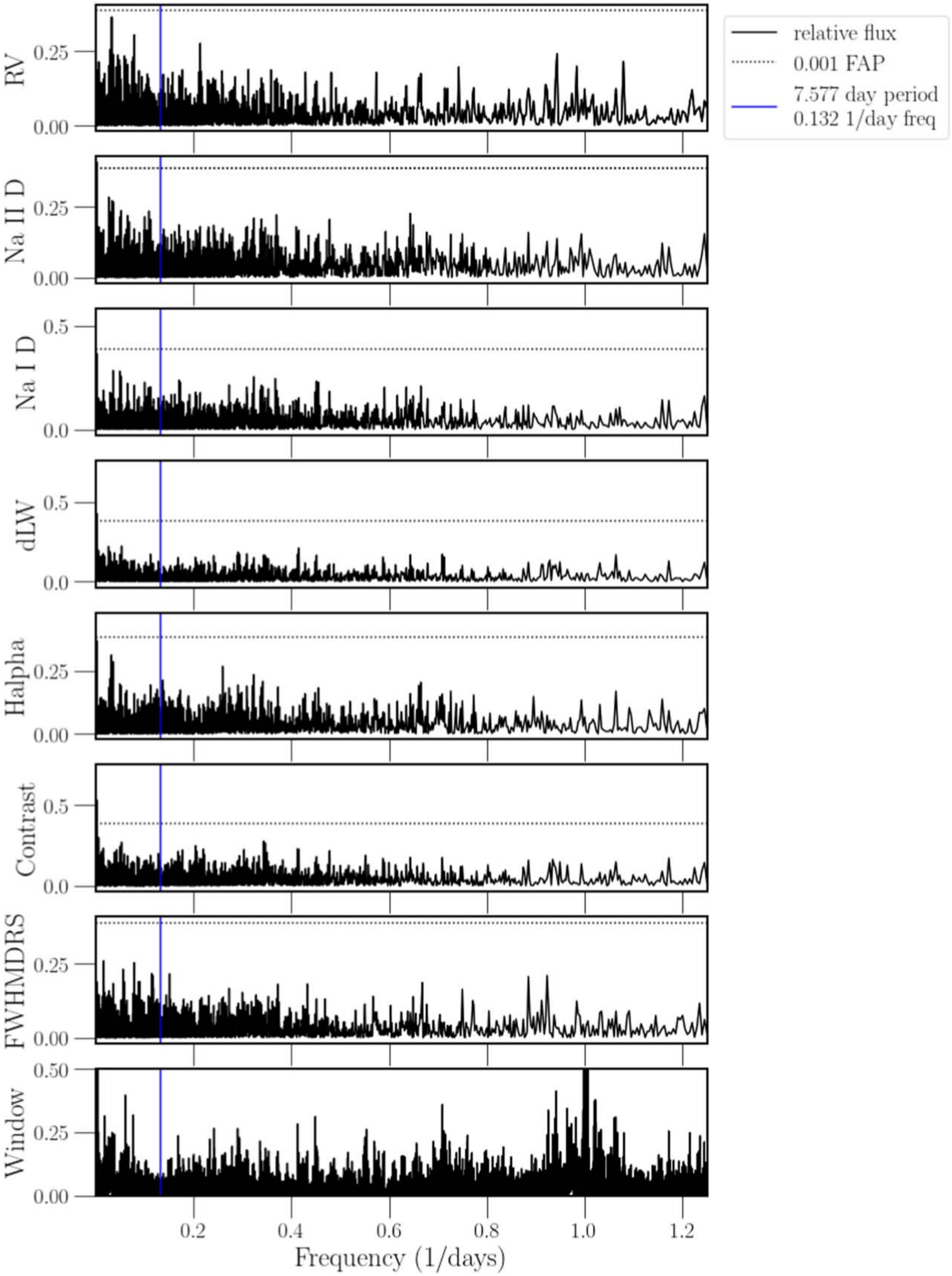

**Figure B8.** Periodograms of the RVs and activity indicators as well as the window function for the HARPS data set. The lack of significant periodicities (with a false-alarm probability less than 0.1%) in these data speaks to the low amplitude of the stellar-rotation-induced activity as well as the lack of planetary signal in the radial velocities. The period of the transiting planet is marked with a blue vertical line in all subplots.

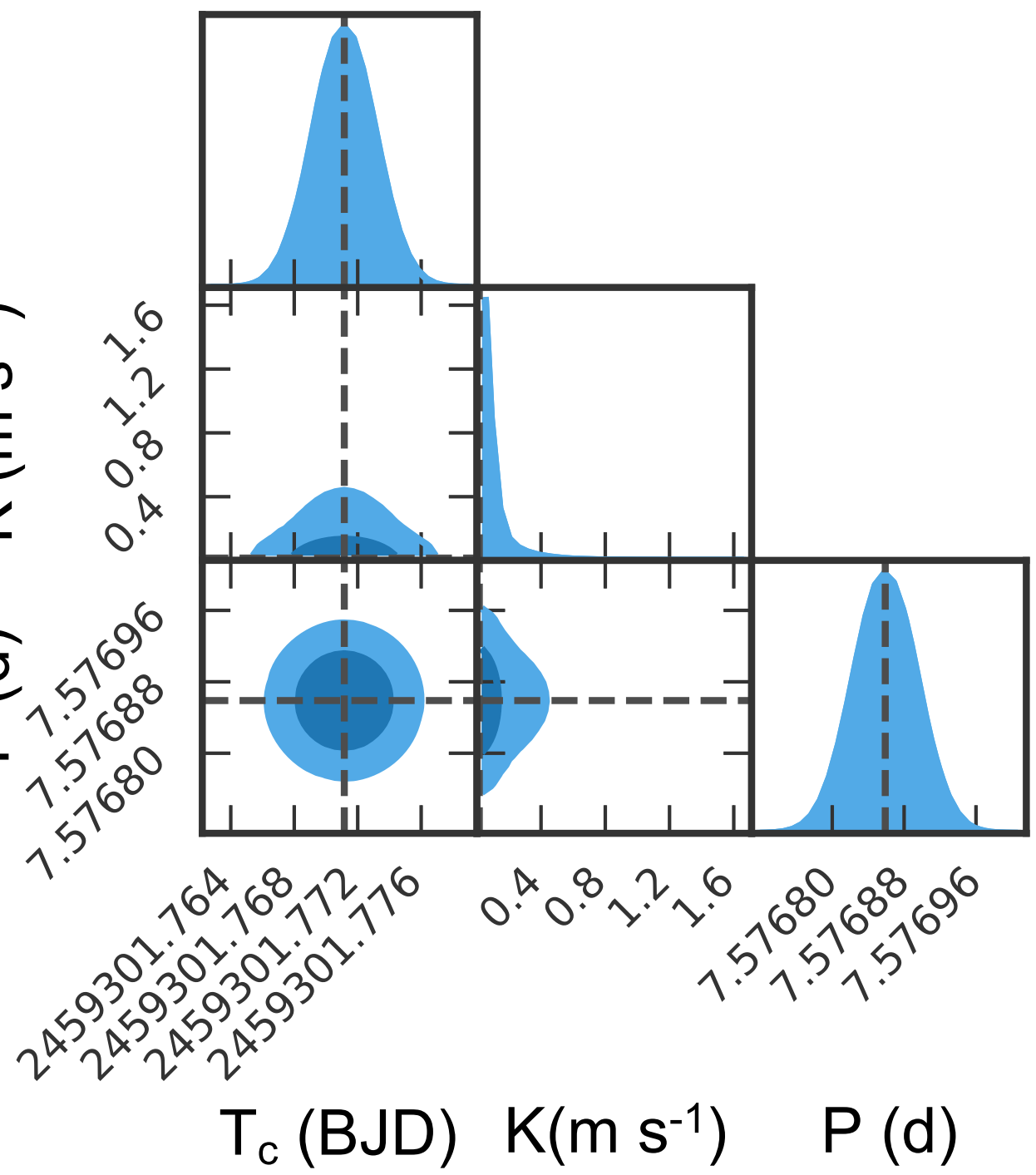

**Figure B9.** Corner plots for RV data fit with a 7.57 day period Keplerian. These parameters were fit simultaneously with the multidimensional, quasiperiodic GP, but corner plots are presented separately for readability.

## ORCID iDs

Victoria DiTomasso https://orcid.org/0000-0003-0741-7661
Mercedes López-Morales https://orcid.org/0000-0003-3204-8183
Sarah Peacock https://orcid.org/0000-0002-1046-025X
Luca Malavolta https://orcid.org/0000-0002-6492-2085
James Kirk https://orcid.org/0000-0002-4207-6615
Kevin B. Stevenson https://orcid.org/0000-0002-7352-7941
Guangwei Fu https://orcid.org/0000-0002-3263-2251
Jacob Lustig-Yaeger https://orcid.org/0000-0002-0746-1980